\documentclass[proof]{WileyASNA-v1}

\articletype{Article Type}%

\received{}
\revised{}
\accepted{}

\begin{document}

\title{Do Hertzsprung-Gap Stars Show Any Chemical Anomaly?}

\author[1,2]{Yoichi Takeda*}
\author[3,4]{Gwanghui Jeong}
\author[3,4]{Inwoo Han}

\authormark{Y. TAKEDA \textsc{et al.}}

\address[1]{\orgname{National Astronomical Observatory of Japan}, 
\orgaddress{\state{2-21-1 Osawa, Mitaka, Tokyo 181-8588}, \country{Japan}}}

\address[2]{\orgname{SOKENDAI, The Graduate University for Advanced Studies}, 
\orgaddress{\state{2-21-1 Osawa, Mitaka, Tokyo 181-8588}, \country{Japan}}}

\address[3]{\orgname{Korea Astronomy and Space Science Institute}, 
\orgaddress{\state{776, Daedeokdae-ro, Youseong-gu, Daejeon 34055}, \country{Korea}}}

\address[4]{\orgname{Korea University of Science and Technology}, 
\orgaddress{\state{217, Gajeong-ro Yuseong-gu, Daejeon 34113}, \country{Korea}}}

\corres{*Y. Takeda. \email{takeda.yoichi@nao.ac.jp}}


\abstract{
With an aim to investigate how the surface abundances of intermediate-mass stars 
off the main sequence (evolving toward the red-giant stage) are affected 
by the evolution-induced envelope mixing, we spectroscopically 
determined the abundances of Li, C, N, O, and Na for selected 62 late A through 
G subgiants, giants, and supergiants, which are often called ``Hertzsprung-gap stars,'' 
by applying the synthetic spectrum-fitting technique to Li~{\sc i} 6708, C~{\sc i} 5380, 
N~{\sc i} 7460, O~{\sc i} 6156--8, and Na~{\sc i} 6161 lines.
A substantially large star-to-star dispersion ($\gtrsim 2$~dex) was confirmed for the Li 
abundances, indicating that this vulnerable element can either suffer significant 
depletion before the red-giant stage or almost retain the primordial composition. 
Regarding C, N, O, and Na possibly altered by dredge-up of nuclear-processed 
products, their abundances turned out to show considerable scatter. 
This suggests that these abundance results are likely to suffer 
appreciable uncertainties, the reason for which is not clear but might be due to 
some kind of inadequate modeling for the atmospheric structure.  
Yet, paying attention to the fact that the relative abundance ratios between C, N, 
and O should be more reliable (because systematic errors may be canceled as 
lines of similar properties are used for these species), 
we could confirm a positive correlation between 
[O/C] (ranging from $\sim 0$ to $\sim +0.5$~dex) and [N/C] (showing a larger 
spread from $\sim 0$ to $\sim +1$~dex), which is reasonably consistent with 
the theoretical prediction.
This observational detection of C deficiency as well as N enrichment in our 
program stars manifestly indicates that the dredge-up of H-burning product 
can take place before entering the red-giant stage, with its extent differing 
from star to star.
}




\maketitle

\footnotetext{\textbf{Abbreviations:} LTE, local thermodynamic equilibrium}

\section{Introduction}

Stars leave off the main sequence after the hydrogen fuel in the core has been 
exhausted and evolve toward the red giant stage with a progressive lowering of 
the surface temperature; i.e., from left to right on the Hertzsprung--Russell 
(HR) diagram. Since the deepening of convention zone takes place during this 
course, some H-burning product in the core may be salvaged to the surface
if the envelope mixing penetrates sufficiently deep, by which characteristic 
chemical anomalies would be observed.  Actually, such signs of abundance 
peculiarities (e.g., C deficit or Na enrichment as a result of contamination by 
CN- or NeNa-cycle product) are actually observed in red giant stars 
(see, e.g., Takeda et al. 2015 and the references therein).

However, it has not yet been well understood how much and when such mixing-induced 
dredge-up actually occurs. According to the theoretical prediction considering 
only the classical convective mixing (canonical theory of first dredge-up),
observable abundance anomalies do not appear until a star has entered the red 
giant phase with sufficiently low $T_{\rm eff}$ ($\lesssim 5000$~K).
Yet, such a conventional theory does not seem to be sufficient and other type 
of additional mixing mechanism is likely to exist in the envelope of actual stars.  
For example, the mixing induced by rotation (e.g., meridional circulation or 
rotational shear instability) may contribute to a significant dredge-up,
which appears to take place already in the main-sequence phase for the case of 
rapidly-rotating B-type stars as seen from their surface He enrichment 
(Lyubimkov et al. 2004). Therefore, recent extensive calculations such as 
done by Lagarde et al. (2012) tend to make practice of including non-canonical 
mixing (rotational as well as thermohaline mixing) in simulating the surface 
abundance changes during the course of stellar evolution.

Even so, we are not sure whether such a more sophisticated modeling with
non-canonical mixing is really superior to the simple classical theory. 
For example, Anthony-Twarog et al. (2018) reported based on their Li abundance 
study of 287 low-to-intermediate mass stars in the open cluster NGC~2506 that 
the observed abundance trend agrees with the prediction from theoretical models 
including rotational+thermohaline mixing. 
On the other hand, Smiljanic et al. (2018) argued in their spectroscopic 
investigation on the C, N, O, Na, and Al abundances of 20 intermediate-mass 
red giants in 10 open clusters that models with rotational mixing tend to 
overestimate the mixing effects and thus not preferable. 

Given the situation being still unsettled as such, it is important in the 
first place to accumulate more observational data concerning the abundances 
of light elements (possibly influenced by mixing) for as many evolved stars 
as possible. Generally speaking, however, previous abundance studies in this 
field tended to focus rather on well-evolved low/intermediate-mass red giants 
of late G or K type, which were B--A--F stars when they were on the main sequence. 
If we are to investigate how the surface abundances undergo changes in the course 
of stellar evolution off the main sequence, it is necessary to pay attention 
also to stars on the mid-way between early-type dwarfs and late-type giants.
Such stars (typically late A through early G giants or supergiants) 
are generally few in number, because they are evolving quite rapidly towards 
the right (cooler) direction on the near-horizontal tracks and thus the probability 
of being found by us small.  Accordingly, we can recognize a void-like region 
of low star density in the HR diagram, which is occasionally called ``Hertzsprung gap.''

Some number of studies regarding the chemical abundances of light elements 
in Hertzsprung-gap stars have been published so far, which we briefly summarize 
as follows (though not meant to be complete): 
Luck \& Lambert (1985) discussed the CNO abundances of F-type supergiants and 
Cepheids in connection with the nature of envelope mixing. 
Takeda \& Takada-Hidai (1994) investigated the behavior of Na abundances in 
A--F supergiants (including some Cepheids). 
Barbuy et al. (1996) determined the CNO abundances of 9 yellow F-type supergiants 
to see if any mixing-related anomaly exists. 
Successively, Smiljanic et al. (2006) reported the CNO abundances of 19 late A 
through early K giants and supergiants. 
Lyubimkov et al. (2011) determined the N abundances of 30 A- and F-type supergiants
and discussed the nature of N-enrichment process.
Takeda et al. (2013) carried out a comprehensive analysis on the C, N, O, and Na
abundances of 12 Cepheid variables.
Adamczak \& Lambert (2014) examined the C and O abundances for 188 stars across 
the Hertzsprung gap, which is probably the most extensive investigation as far as
the number of sample stars is concerned.
Molina \& Rivera (2016) reported the chemical abundances of many elements (including
C, N, O, and Na) in 4 A--F supergiants.

In spite of these investigations, however, the picture of evolution-induced 
chemical anomaly in this group of stars has not been clarified yet.
Besides, targets in these past studies, except for Adamczak \& Lambert 
(2014), appear to be biased toward comparatively slow rotators, despite rapid 
rotators are commonly included in this group of stars (presumably because of the 
growing difficulty in abundance determinations).

Given this circumstance, we decided to challenge this task, 
taking advantage of the results and experiences in our recent studies:\\
--- Regarding the observational data, the high-dispersion spectra of 75 evolved 
A-, F-, and G-type stars in the Hertzsprung-gap region are already available, 
which we used for investigating the luminosity effect of O~{\sc i} 7771--5 
triplet lines (Takeda et al. 2018a; hereinafter referred to as Paper~I).
Besides, the atmospheric parameters of these stars (including the microturbulence 
which is not easy to determine for broad-line stars) have already been established 
in Paper~I.\\
--- As to the surface chemical composition of red giants (into which
Hertzsprung-gap stars will further evolve sooner or later), Takeda et al. (2015) 
reported the abundances of C, O, and Na for 239 late G through early K giants.
Likewise, Takeda et al. (2013) investigated the C, N, O, and Na abundances of
12 Cepheid variables, which may also be used for comparison.\\
--- On the other hand, surface C, N, and O abundances of late B through early F dwarfs 
(from which $\sim$~1.5--5~$M_{\odot}$ stars currently in the Hertzsprung gap 
have evolved) have recently been investigated in detail by Takeda et al. 
(2018c; hereinafter referred to as Paper~II).\\
--- Since these previous studies made use of the spectrum synthesis technique, 
which is indispensable for deriving the elemental abundances of rapid rotators 
often included in Hertzsprung-gap stars, we may be able to compare the abundances
of different star groups in a consistent manner by applying the same method of analysis. 

Accordingly, we investigated in this study the abundances of representative light 
elements (Li, C, N, O, and Na, which may be affected by evolution-induced mixing) 
for selected 62 Hertzsprung-gap stars (late A through G subgiants, giants, 
and supergiants) by making use of the observational data adopted in Paper~I. 
Here, we particularly intended to examine the following points of interest.
\begin{itemize}
\item
Do these stars currently evolving across the Hertzsprung gap show abundance 
anomalies typically seen in red giants or Cepheids (e.g., underabundance in C, 
overabundance in N or Na)? If really observed, is there any meaningful dependence 
upon the stellar parameters (e.g., $T_{\rm eff}$ or $M$)?
\item
How are the abundances of these evolved stars compared with low-mass (FGK) dwarfs
covering wide range of stellar ages? Meanwhile, do they show any relation with 
the C, N, and O deficiencies observed in A-type dwarfs (most likely caused 
by the diffusion process and confined only to the surface layer)? 
\end{itemize}

\section{Program stars and their parameters}

Among the 75 stars investigated in Paper~I, the effective temperatures of early 
A-type supergiants tend to suffer appreciably larger ambiguities (cf. Fig.~2a therein), 
mainly because of the considerable interstellar extinction (due to their distant 
nature with low galactic latitude). Accordingly, we decided to discard 13 stars 
(those with $T_{\rm eff} > 8500$~K or classified as early A supergiants), 
which eventually resulted in 62 program stars, as listed in Table~1. 
See Sect.~2 in Paper~I for the description of the observational data of these 
targets, which were obtained with the echelle spectrograph attached to
the 1.8~m reflector at Bohyunsan Astronomical Observatory.
 
Regarding $T_{\rm eff}$ (effective temperature; from $B-V$ colors) and 
$\log g$ (surface gravity; from luminosity with the help of evolutionary tracks), 
we used the same values as derived in Paper~I (cf. Sect.~3 therein).
As to the microturbulence ($v_{\rm t}$), we adopted the values determined
in Paper~I by requiring the abundance consistency between O~{\sc i} 7771--5 
and O~{\sc i} 6156--8 lines (denoted as $\xi_{\rm a}$; cf. Sect.~6 therein)
wherever possible. In case that $\xi_{\rm a}$ could not be determined in Paper~I, 
we used the alternative microturbulence ($\xi_{\rm p}$) determined from the 
line profile of O~{\sc i} 7771--5 triplet (cf. Sect.~5 therein). 
The model atmosphere assigned to each star is described in Sect.~4 of Paper~I. 

The $\log L$ vs. $\log T_{\rm eff}$ plots of our 62 Hertzsprung-gap stars 
are shown in Fig.~1, where the relevant targets of different star groups 
(red giants, Cepheids, main-sequence stars) studied in our previous papers
are also depicted.
Similarly, $\log g$, $v_{\rm t}$, and $v_{\rm e}\sin i$ (projected rotational
velocity determined from 6145--6166~\AA\ fitting described in Sect.~3.1) 
for each star are plotted against $T_{\rm eff}$ and $M$ (stellar mass) in Fig.~2. 
We can see from these figures that our sample stars cover the parameter 
ranges of  8000~K~$\gtrsim T_{\rm eff} \gtrsim 5000$~K,
1.8~$\lesssim \log g \lesssim$~3.7, $1.5M_{\odot} \lesssim M \lesssim 8M_{\odot}$, 
0~km~s$^{-1} \lesssim v_{\rm t} \lesssim 10$~km~s$^{-1}$, and 
0~km~s$^{-1} \lesssim v_{\rm e}\sin i \lesssim 150$~km~s$^{-1}$.

As for the error bars ($\sigma$) attached to the data points in Fig.~1 
and Fig.~2, $\sigma(T_{\rm eff})$'s are due to ambiguities of interstellar
reddening, $\sigma(\log L)$'s are evaluated by combining the uncertainties
of interstellar extinction and of Hipparcos parallax, $\sigma(M)$'s
are due to ambiguities in $L (\propto M^{4})$, and $\sigma(\log g)$'s
are estimated from $\sigma(T_{\rm eff})$ and $\sigma(L)$ (where we 
formally assume that both are independent and that random errors in $M$
are comparatively insignificant and negligible).
The adopted stellar parameters for each program star are summarized 
in Table~1 (and in ``tableE.dat'' of the online material where their 
errors are also given). 

\section{Abundance determinations}
 
Given that our task is to study the surface abundances of Li, C, N, O, and Na 
for 62 Hertzsprung-gap stars, we invoke Li~{\sc i} 6708, C~{\sc i} 5380, 
O~{\sc i} 6156--8, N~{\sc i} 7468, and Na~{\sc i} 6161 lines as in 
Takeda \& Tajitsu (2017) (for Li, C, O, Na) and Paper~II (for C, N,and O).
The determination procedures of abundances and related quantities (e.g., 
non-LTE correction, uncertainties due to ambiguities of atmospheric parameters) 
are essentially the same as described in these papers, which consist 
of two consecutive steps.

\subsection{Synthetic spectrum fitting} 

The first step is to find the solutions for the abundances of relevant elements 
($A_{1}, A_{2}, \ldots$), projected rotational velocity ($v_{\rm e}\sin i$), 
and radial velocity ($V_{\rm rad}$) by requiring the
best fit (minimizing $O-C$ residuals) between theoretical and 
observed spectra, while applying the automatic fitting algorithm 
(Takeda 1995a). Four wavelength regions were selected for this purpose:
(i) 6702--6714~\AA\ region (for Li), 
(ii) 5370--5390~\AA\ region (for C), 
(iii) 7457--7472~\AA\ region (for N), and
(iv) 6145--6166~\AA\ region (for O, Na). 
More information about this fitting analysis (varied elemental abundances, 
used data of atomic lines) is summarized in Table~2.
How the theoretical spectrum for the converged solutions fits well 
with the observed spectrum is displayed in Fig.~3--6 for each region. 
The $v_{\rm e}\sin i$ values\footnote{This $v_{\rm e}\sin i$ is the same
as what we referred to as $v_{\rm M}$ in Paper~I. It should be 
kept in mind that we assumed only the rotational broadening (with the 
limb-darkening coefficient of $\epsilon = 0.5$) as the macrobroadening 
function to be convolved with the intrinsic theoretical line profiles.
Accordingly, $v_{\rm e}\sin i$ values for sharp-line cases 
(e.g., $v_{\rm e} \sin i \lesssim$~10~km~s$^{-1}$) 
should be regarded rather as upper limits because the effects of 
instrumental broadening and macroturbulence were neglected.
} resulting from the fitting of 6145--6166~\AA\ region are presented 
in Table~1. We also adopted the solution of Fe abundance derived from 
the fitting of 6146--6163~\AA\ region as the metallicity of each star 
(given as [Fe/H] in Table~1).  

\subsection{Abundances from equivalent widths} 

As the second step, with the help of Kurucz's (1993) WIDTH9 program 
(which had been considerably modified in various respects; e.g., 
inclusion of non-LTE effects, treatment of total equivalent width for 
multi-component lines; etc.), we computed the equivalent widths ($W$) 
of the representative lines ``inversely'' from the abundance solutions
(resulting from spectrum synthesis) along with the adopted atmospheric 
model/parameters; i.e., $W_{6708}$ (for Li~{\sc i} 6708), $W_{5380}$ 
(for C~{\sc i} 5380), $W_{6156-8}$ (for O~{\sc i} 6156--8), 
$W_{7468}$ (for N~{\sc i} 7468), and $W_{6161}$ (for Na~{\sc i} 6161),
which are easier to handle in practice (e.g., for estimating the
uncertainty due to errors in atmospheric parameters).
The adopted atomic data for these lines are summarized in Table~3.  
We then analyzed such derived $W$ values by using WIDTH9 to determine 
$A^{\rm N}$ (NLTE abundance) and $A^{\rm L}$ (LTE abundance),\footnote{
$A_{\rm X}$ is the logarithmic number abundance of element X expressed
in the usual normalization of $A_{\rm H} = 12$; i.e., 
$A_{\rm X} = \log (N_{\rm X}/N_{\rm H}) + 12$.} 
from which the NLTE correction $\Delta (\equiv A^{\rm N} - A^{\rm L})$
was further derived. We adopted Procyon as the standard star of 
abundance reference (except for Li, which should be compared with 
the solar system abundance of 3.31 as done in Takeda \& Tajitsu 2017), 
because it is known to have practically  
the same abundance as the Sun. We thus define the relative abundance as 
[X/H] $\equiv$ $A_{\rm X}^{\rm N}$(star) $-$ $A_{\rm X}^{\rm N}$(Procyon) 
(X = C, N, O, and Na). 
The references abundances of $A_{\rm X}^{\rm N}$(Procyon) (to be determined
in the same manner and with the same atomic data as adopted in this study) 
are 8.70 (C), 8.10 (N), 8.83 (O), 6.29 (Na), and 7.47 (Fe), where
those of C, N, O, and Fe were taken from Paper~II and that of Na
was newly determined in this study. The resulting values of $A^{\rm N}$(Li), 
[C/H], [N/H], [O/H], [Na/H] are given in Table~1 (more complete results
including $W$ and $\Delta$ are separately presented in ``tableE.dat''). 
Figs. 7(C), 8(C), 9(N), 10(O), and 11(Na) graphically show the equivalent 
width ($W$), non-LTE correction ($\Delta$), non-LTE abundance ($A^{\rm N}$), 
and abundance variations in response to parameter changes (see the 
following Sect.~3.3), as functions of $T_{\rm eff}$. 

\subsection{Error estimation}

In order to evaluate the abundance errors caused by uncertainties
in atmospheric parameters, we estimated six kinds of abundance variations
($\delta_{T+}$, $\delta_{T-}$, $\delta_{g+}$, $\delta_{g-}$, 
$\delta_{v+}$, and $\delta_{v-}$) for $A^{\rm N}$ by repeating the 
analysis on the $W$ values while 
perturbing the standard atmospheric parameters interchangeably by 
$\pm\sigma(T_{\rm eff)}$ (cf. Sect.~2), $\pm\sigma(\log g)$ (cf. Sect.~2), 
and $\pm\sigma(v_{\rm t})$ (which we tentatively assumed 
$\pm$max$[1.0, 0.3 v_{\rm t}]$~km~s$^{-1}$ by consulting Fig.~10d in Paper~I). 
Finally, the root-sum-square of these perturbations,
$\delta_{Tgv} \equiv (\delta_{T}^{2} + \delta_{g}^{2} + \delta_{v}^{2})^{1/2}$, 
were regarded as the abundance uncertainties due to combined errors in 
$T_{\rm eff}$, $\log g$, and $v_{\rm t}$, 
where $\delta_{T}$, $\delta_{g}$, and $\delta_{\xi}$ are defined as
$\delta_{T} \equiv (|\delta_{T+}| + |\delta_{T-}|)/2$, 
$\delta_{g} \equiv (|\delta_{g+}| + |\delta_{g-}|)/2$, 
and $\delta_{v} \equiv (|\delta_{v+}| + |\delta_{v-}|)/2$,
respectively. 
These $\delta_{T\pm}$, $\delta_{g\pm}$, and $\delta_{v\pm}$ are plotted
against $T_{\rm eff}$ in panels (d), (e), and (f) of Figs.~7--11,
from which can generally state that $\delta_{T\pm}$ is most significant 
(which can be as large as $\sim 0.2$~dex or even more), while the other 
two are of comparatively minor importance. 

We also evaluated errors in the equivalent width ($\delta W$) by assuming 
the typical S/N of $\sim 200$, from which the corresponding abundance 
uncertainties ($\delta_{W}$) were derived as done in Paper~II (see 
Sect.~4.3 therein).
Since $\delta W$ is quite small (typically several tenths to a few m\AA\
depending on $v_{\rm e}\sin i$), $\delta_{W}$ is only a few hundredths dex 
in most cases, except for the case of very weak lines where $\delta_{W}$ 
can be appreciably large as much as several tenths dex. 
Since the abundance results are regarded as unreliable if $W$ is smaller 
than 3$\delta_{W}$ (cf. Sect.~4.3 in Paper~II), the relevant solutions 
of 3 stars (Li), 3 stars (N), 2 stars (O), and 1 star (Na) satisfying 
this criterion were discarded (these rejected data points are indicated 
by open circles in panels (a) and (c) of Figs. 7, 9, 10, and 11). 

Finally, combining $\delta_{Tgv}$ and $\delta_{W}$, we evaluated
the total error as $\delta_{TgvW} \equiv (\delta_{Tgv}^{2} + \delta_{W}^{2})^{1/2}$,
which are shown as error bars attached to the non-LTE abundances in panel (c)
of Figs.~7--11, though $\delta_{TgvW}$ is generally dominated by $\delta_{Tgv}$ 
(i.e., $\delta_{TgvW} \simeq \delta_{Tgv}$). 

Meanwhile, we adopted $A$(Fe) derived from the spectrum fitting in the 6145--6166~\AA\ 
region as the representative Fe abundance to obtain [Fe/H]. In this case, evaluation 
of error in $A$(Fe) is not straightforward, because Fe~{\sc i} lines as well as 
Fe~{\sc ii} lines are involved (see Fig.~6). As a tentative solution, postulating
that Fe~{\sc ii} 6149.258 line (which has an appreciable strength in this region)
is most important in determining $A$(Fe), we evaluated $\delta_{TgvW}$ from $W$(6149)
in the same manner as mentioned above.
The $\delta_{TgvW}$ values involved with the abundances of Li, C, N, O, Na, and Fe 
for each star are given in ``tableE.dat.''  

\section{Discussion}

\subsection{Apparent trends of the abundances}

The resulting abundances of Li, C, N, O, Na, and Fe relative to the reference values
(solar system abundance for Li, Procyon abundances for the other elements being 
almost equal to the solar composition) are plotted against $T_{\rm eff}$ and $M$
in Fig.~12. The mutual correlations between these abundances and the 
[N/C] vs. [O/C] relation are shown in Fig.~13, where how the ratios of 
[C/Fe], [N/Fe], [O/Fe], and [Na/Fe] behave with a change of [Fe/H] (along 
with the relevant diagrams for FGK dwarfs and red giants for comparison) 
is also depicted.  

The drastic spread of $A$(Li) is immediately noticeable in Fig.~12a,a'.
Since Li abundances could be established only $\sim 40$\% (25 out of 62) of 
the program stars, surface Li in the remaining (more than half) stars must also 
be substantially depleted ($A$(Li)$ < 1$) because the upper limits (corresponding 
to $W_{6708}$ from a few tenths m\AA\ to a few m\AA) is $A$(Li)~$\sim$~0--1. 
Accordingly, the overall star-to-star dispersion of $A$(Li) should be $\gtrsim 2$~dex 
or even more, which means that the Li-dilution process in the envelope of 
Hertzsprung-gap stars is considerably case-dependent. Considering the
vulnerability of this element (which is quickly destroyed at comparatively 
low temperature of $T \sim 2.5\times 10^{6}$~K), this diversity may be 
understandable. 

However, it was rather unexpected that appreciably large scatter is seen in  
the diagrams involving [C/H], [N/H], [O/H], [Na/H], and [Fe/H] (Fig.~12, Fig.~13),
which apparently contrasts with the case of red giants (e.g., Takeda et al. 2015). 
Actually, the spread in these [X/H] values extends typically to $\sim \pm$0.5--0.6~dex
or even more (e.g., the case of [N/H]). This can hardly be regarded 
as the metallicity effect or some chemical anomaly effect, because
their mutual correlation is not good (near-random distribution around 
[X/H]~$\sim 0$ might rather be a better description) as seen from Fig.~13, 
though reasonable correlation is observed in some cases (e.g., 
[C/H] and [O/H] in Fig.~13f).  
We thus can not help considering that significant errors are involved 
in our abundance results, possibly larger than the error bars 
in each figure panel ($\delta_{TgvW}$; cf. Sect.~3.3).
It is likely that our adopted atmospheric models were not sufficiently
adequate, for which we may think of two factors (errors in the atmospheric 
parameters, impact of chromospheric activity) as ponderable possibilities,
as separately discussed in Appendix~A.

\subsection{Implication of evolution-related anomalies}

As such, we must realize that the abundances 
derived in Sect.~3 are likely to contain significant errors due to imperfections 
of our analysis, and thus their face values should not be blindly trusted.
Yet, we can try to extract as much information regarding 
the nature of envelope mixing as possible based on the resulting abundances.
 
We have already seen that the surface Li abundances of our program stars
are very diversified (spanning a range of $\gtrsim 2$~dex) from the near-primordial 
abundance of $A$(Li)$\sim 3$ down to a very depleted level of $A$(Li)$\lesssim 1$ 
(cf. Fig.~12a,a'). This means that, while some stars have not experienced 
any substantial mixing (because Li is retained), efficient mixing-induced dilution 
takes place for other stars. It is worth noting that considerably Li-depleted stars
do exist at 7000~K~$\gtrsim T_{\rm eff} \gtrsim$~6000~K. This suggests that some kind of 
non-canonical mixing (e.g., rotational or thermohaline mixing) is required for 
such stars, because depletion of surface Li begins after a star has become 
sufficiently cool at 5500--5000~K~$\gtrsim T_{\rm eff}$ according to the conventional
theory including only the standard mixing (cf. Fig.~7a in Takeda et al. 2018b).
Besides, those stars almost retaining the original Li contents tend to rotate
comparatively rapidly, suggesting that Li depletion process operates more 
efficiently for slower rotators.
 
Then, what about the dredge-up of nuclear-processed products? 
Unfortunately, as mentioned in Sect.~4.1, we can not place much confidence on 
the apparent [C/H], [N/H], [O/H], and [Na/H] values themselves, because of their 
considerable dispersion which must have masked meaningful trends (if any exists). 
However, we would hope that the {\it relative} difference between the abundances of 
the similar type of species may still be relied upon, because the systematic 
errors would act on the same direction. Accordingly, it is worth paying attention 
to the relative ratios of C, N, and O abundances, because they were derived from 
similar high-excitation lines of similar dominant-population species 
(C~{\sc i}, N~{\sc i}, O~{\sc i}). 

We thus decided to focus on two abundance ratios: [N/C] and [O/C].
The [N/H] vs. [C/H] diagram (Fig~13e) indicates that [N/H] values are very diversified,
leading to a considerably large spread of [N/C] ($0 \lesssim$~[N/C]~$\lesssim 1$). 
Meanwhile, Fig.~13f shows that [C/H] and [O/H] correlate with each other 
([O/H]$-$[C/H] $\sim 0.3$ with a dispersion of $\sim \pm$0.2--0.3~dex),
which means that the span of [O/C] is rather moderate ($0 \lesssim$~[O/C]~$\lesssim 0.5$). 
Actually, we can confirm from the [N/C] vs. [O/C] diagram (Fig.~13h) that
these two show a positive correlation over the ranges mentioned above.

Then, recalling that oxygen is least affected among these
three elements (being hardly changed or only slightly decreased; see, e.g., 
Fig.~11 in Takeda et al. 2015) and regarded as nearly normal, we can see 
the tendency of C deficiency as well as N enrichment to hold in our sample stars,
which is most naturally interpreted as due to the contamination of CN-cycled material
salvaged from the deep H-burning region due to an efficient mixing.\footnote{
The chemical anomalies of CNO seen in most of the A-type dwarfs are characterized by
their underabundances with an inequality tendency of [C/H]~$\lesssim$~[N/H]~$\lesssim$~[O/H]~$\lesssim 0$ 
(cf. Paper II). It is unlikely that the CNO abundance patterns of our program stars 
under question are associated with those of A-type main sequence stars, 
because the observed trend of of [N/C]~$>$[O/C]~$(\gtrsim 0)$ is totally 
incompatible with such an inequality relation.} 
Actually, the observed [N/C] vs. [O/C] correlation is reasonably consistent 
with the theoretically predicted locus (cf. the thick line in Fig.~13h) computed by
Lagarde et al. (2012).

It is difficult, however, to include sodium in this discussion, because 
possible systematic errors involved in the Na abundances (derived from Na~{\sc i}; minor 
population species of low ionization potential) are considered to act in a markedly 
different sense than those of CNO. Still, we can recognize in Fig~12e,e' that 
the [Na/H] values of higher $T_{\rm eff}$ stars (presumably less affected by 
activity-related errors) tend to be positive and to slightly increase with $M$ at 
$4 M_{\odot} \lesssim M \lesssim 8 M_{\odot}$, such as reported by Takeda \& Takada-Hidai (1994)
for A--F supergiants of $M \gtrsim 10 M_{\odot}$. This may indicate a sign of $M$-dependent 
dredge-up of NeNa-cycle product.

Taking these results into consideration, we may conclude that (i) evolution-induced 
surface abundance anomalies do exist in (at least a significant fraction of) our sample 
stars across the Hertzsprung gap, (ii) appreciable dredge-up of H-burning product 
must take place before entering the red-giant stage in those stars, and (iii)
whether and how much a star experiences such a dredge-up is considerably case-dependent 
(as seen from the large diversity of [N/C]).   

\subsection{Comparison with other studies}

Finally, we briefly comment on how the results derived from our analysis of 62 program 
stars are compared with those reported by previous investigators. Generally speaking,
our conclusion appears to be compatible with most of the relevant past studies, 
at least in the qualitative sense.

Regarding lithium, our results, that the surface Li abundances show 
a very large diversity (ranging from the near-primordial value of $A$(Li)$\sim 3$ 
down to a considerably depleted level of $\lesssim 1$) and that rapid rotators tend to 
retain the original composition without appreciable depletion, are in good agreement 
with what de Laverny et al. (2003) concluded in their study of 54 giants across 
the Hertzsprung gap (cf. their Fig.~3). 

Likewise, our conclusion concerning the CNO abundances (Hertzsprung-gap stars generally 
show signs of chemical anomalies caused by the dredge-up of H-burning products 
though the extents widely differ from star to star) is reasonably consistent 
with several related studies mentioned in Sect.~1, most of which similarly reported 
the existence of characteristic chemical signature (such as the deficiency of 
C and/or enrichment of N or Na) indicating that the nuclear-processed material 
had been more or less salvaged from the inner H-burning region.

However, one exception is the recent extensive spectroscopic study of 188 
Hertzsprung-gap stars conducted by Adamczak \& Lambert (2014), who concluded
that their C and O abundances were almost similar to those of low mass dwarfs
and no indication of significant mixing-induced abundance changes was found.
Therefore, their conclusion is in conflict with ours as well as those 
of other published work. Since 10 stars among their sample (HD~26553, 72779, 
82543, 88759, 100418, 111812, 151070, 164136, 188650, and 201078) 
are common to our program stars, their stellar parameters as well as 
the Fe, C, and O abundances are compared with those derived by us in Fig.~14,
from which we can see the following characteristics:
Regarding the fundamental and atmospheric parameters, $T_{\rm eff}$ and $\log g$ 
are more or less consistent, whereas $M$ as well as $v_{\rm e}\sin i$ are 
in fairly good agreement. 
However, $v_{\rm t}$ is in serious disagreement (cf. Fig.~14c) in the sense 
that our values are diversified in the range of $\sim $~2--10~km~s$^{-1}$ while theirs
show a much smaller spread around $\sim 2$~km~s$^{-1}$ (presumably because they 
assumed $v_{\rm t} = 1.8 \pm 0.35$~km~s$^{-1}$ in cases where this parameter 
could not be determined). Since this Fig.~14c is apparently similar to
Fig.~A3c in Appendix A.1, where we point out the possibility of overestimation 
for our prominently large $v_{\rm t}$ values around $T_{\rm eff} \sim 6000$~K 
(cf. Fig.~2b), the same explanation may as well be applied for this discrepancy. 
As to the relative abundances of 
[Fe/H], [C/H], and [O/H], Figs.~14f--14h indicate that our results have little
correspondence with theirs (i.e., the former tend to spread over a much wider 
range than the latter). Although the details of their analysis (e.g., equivalent 
widths, line-by-line abundances) are not published, they seem to have employed 
the high-excitation C~{\sc i} lines at 5052, 5380, and 6587~\AA\
for C abundance determination, while the forbidden [O~{\sc i}] lines 
of low excitation at 6300, 6363, and 5577~\AA\ were used along 
with the strong high-excitation O~{\sc i} 7771--5 triplet lines (appreciably 
affected by the non-LTE effect as well as by a choice of microturbulence $v_{\rm t}$).
It thus appears somewhat questionable whether the trend of C and O abundances 
(and the C/O ratios) could be reliably derived by combining the results from
such lines of considerably different characteristics. Besides, in our opinion, 
a weakpoint in their study is that they did not take into account N, which plays 
an important role in discussing the chemical anomaly due to mixing of H-burning product 
because its abundance change is obviously large (compared to that of C or O).  
Accordingly, it seems still premature to regard the conclusion of 
Adamczak \& Lambert (2014) as being established, which would need to be checked 
by follow-up investigations. 
 
\section{Conclusion}

Those evolved intermediate-mass stars, which have left the main sequence and 
are currently on the way toward the red-giant stage (from left to right on 
the HR diagram), are often called ``Hertzsprung-gap stars.'' 
In this study, we intended to examine whether and how their surface abundances 
of light elements show characteristic anomalies caused by evolution-induced 
mixing in the envelope.

Toward this aim, we determined the abundances of Li, C, N, O, and Na for 
selected 62 late A through G subgiants, giants, and supergiants 
by applying the spectrum fitting technique to Li~{\sc i} 6708, C~{\sc i} 5380, 
N~{\sc i} 7460, O~{\sc i} 6156--8, and Na~{\sc i} 6161 lines, based on the 
high-dispersion spectra obtained with the 1.8~m reflector at Bohyunsan 
Astronomical Observatory.

We confirmed a substantially large star-to-star dispersion ($\gtrsim 2$~dex) in the Li 
abundances, indicating that this vulnerable element can either suffer significant 
depletion before the red-giant stage or still retain most of the primordial 
composition, and the latter case tends to be seen in rapid rotators. 

Somewhat disappointingly, it turned out difficult to perceive meaningful trends
from the abundances of C, N, O, and Na (possibly altered by the dredge-up of 
H-burning products) by themselves, because of the considerably large abundance 
scatters which can not be real but due to some additional significant errors.  

Despite such an disadvantage of having to deal with the abundance data involving
considerable uncertainties, it may be hoped that relative abundance ratios between 
C, N, and O may still be relied upon, because errors tend to be canceled as 
these abundances were derived from similar high-excitation lines of dominant
species. Following this consideration, we found a positive correlation between 
[O/C] (ranging from $\sim 0$ to $\sim +0.5$~dex) and [N/C] (showing a larger spread 
from $\sim 0$ to $\sim +1$~dex). Moreover, this trend is reasonably consistent 
with the theoretical prediction based on stellar evolution calculations.

This corroborates that the abundance characteristics caused by contamination of 
nuclear-processed material (C deficiency as well as N enrichment) are detected
in the CNO abundances of our program stars. Accordingly, we can conclude that 
the dredge-up of H-burning product can take place before entering the red-giant 
stage, though its extent differing from star to star.

\section*{Acknowledgments}

This research has made use of the SIMBAD database operated at CDS, Strasbourg, 
France. Data reduction and analysis were in part carried out by using the 
common-use data analysis computer system at the Astronomy Data Center (ADC) 
of the National Astronomical Observatory of Japan.



\setcounter{table}{0}
\begin{table*}[h]
\footnotesize
\caption{Stellar parameters and the resulting abundances.}
\begin{center}
\begin{tabular}{crccccccrcccccc}\hline\hline
HD\# & HIP\# & Sp.Type & $\log L$ & $M$ & $T_{\rm eff}$ & $\log g$ & $v_{\rm t}$ & 
$v \sin i$ & [Fe] & $A_{\rm Li}$ & [C] & [N] & [O] & [Na] \\
(1) & (2) & (3) & (4) & (5) & (6) & (7) & (8) & (9) & (10) & (11) & (12) & (13) & (14) & (15) \\
\hline
000571&    841& F2II        & 3.20& 5.42& 6995& 2.30&  5.7&  53.9& $-$0.54& $\cdots$& $-$0.42& $-$0.08& $-$0.41& +0.07\\
006130&   4962& F0II        & 3.52& 6.54& 7616& 2.22&  4.8&  16.2& $-$0.09& $\cdots$& $-$0.30& +0.03& $-$0.34& +0.34\\
006210&   5021& F6V         & 1.45& 1.97& 5983& 3.34&  8.5&  46.1& $-$0.31& $\cdots$& $-$0.10& $-$0.38& $-$0.57& $-$0.24\\
007034&   5544& F0V         & 2.23& 2.97& 7349& 3.10&  4.4& 109.2& $-$0.03& $\cdots$& $-$0.65& +0.58& $-$0.11& $\cdots$\\
008890&  11767& F7:Ib-IIv SB& 3.43& 6.50& 5741& 1.81&  4.4&  12.0& $-$0.14& $\cdots$& $-$0.20& +0.33& +0.09& $-$0.02\\
020902&  15863& F5Ib        & 3.65& 7.35& 6389& 1.83&  5.6&  19.6& $-$0.16& $\cdots$& $-$0.39& +0.28& $-$0.08& +0.05\\
022211&  16695& G0          & 1.63& 2.27& 5720& 3.14&  6.4&  94.8& $-$0.15&  1.93& +0.30& +0.28& $\cdots$& $\cdots$\\
023230&  17529& F5IIvar     & 2.96& 4.71& 6728& 2.42&  4.3&  45.0& $-$0.17&  1.97& $-$0.40& $-$0.03& $-$0.20& $-$0.01\\
025291&  19018& F0II        & 3.76& 7.65& 7677& 2.06&  4.9&   7.6& $-$0.17& $\cdots$& $-$0.27& $-$0.01& $-$0.37& +0.45\\
026553&  19823& A4III       & 3.06& 5.00& 6767& 2.35&  3.2&   6.3& $-$0.58& $\cdots$& $-$0.92& +0.68& +0.26& $-$0.63\\
032655&  23799& F2IIp...    & 2.25& 3.08& 6548& 2.90&  4.5&  29.0& $-$0.10&  1.61& $-$0.11& +0.10& +0.21& $-$0.20\\
038529&  27253& G4V         & 0.79& 1.46& 5320& 3.67&  1.5&   6.6& +0.15& $\cdots$& +0.23& +0.25& +0.47& +0.20\\
038558&  27338& F0III       & 3.33& 5.80& 7579& 2.34&  5.0&  25.4& $-$0.31& $\cdots$& $-$0.21& $-$0.64& $-$0.38& +0.04\\
045412&  30827& F5.5Ibv     & 3.19& 5.60& 5694& 1.98&  4.7&  13.1& $-$0.26& $\cdots$& $-$0.03& +0.80& +0.49& $-$0.08\\
050018&  33041& F2V         & 1.88& 2.51& 6660& 3.21&  2.8& 146.1& $-$0.02& $\cdots$& $-$0.12& +0.63& +0.31& $\cdots$\\
050420&  33269& A9III       & 2.51& 3.56& 7229& 2.87&  4.0&  29.0& $-$0.26& $\cdots$& $-$0.50& $-$0.02& $-$0.26& +0.07\\
057749&  35749& F3IV        & 2.33& 3.21& 6803& 2.90&  3.0&  41.4& +0.04&  1.87& $-$0.31& +0.37& +0.09& +0.14\\
059881&  36641& F0III       & 2.55& 3.63& 7332& 2.86&  4.4&  59.1& $-$0.12& $\cdots$& $-$0.31& +0.03& $-$0.04& $-$0.02\\
072779&  42133& G0III       & 1.97& 2.78& 5596& 2.86& 10.0&  98.6& $-$0.31&  2.92& $\cdots$& $-$0.19& $\cdots$& $-$0.52\\
077601&  44613& F6II-III    & 1.71& 2.30& 6216& 3.22&  6.9& 142.6& +0.08& $\cdots$& $-$0.53& +0.63& +0.05& $\cdots$\\
082210&  46977& G4III-IV    & 1.17& 1.88& 5294& 3.39&  4.8&   8.6& $-$0.41&  1.21& $-$0.54& $\cdots$& +0.12& $-$0.27\\
082543&  46840& F7IV-V      & 1.63& 2.26& 5742& 3.15&  4.9&   6.9& $-$0.26&  1.96& $-$0.30& $-$0.57& +0.19& $-$0.05\\
088759&  50286& F2          & 1.77& 2.36& 6518& 3.25&  7.8&  90.3& $-$0.36&  3.06& $-$0.15& $-$0.41& $-$0.35& $\cdots$\\
100418&  56364& F8/G0Ib/II  & 1.72& 2.39& 5847& 3.12&  2.0&  32.9& $-$0.11& $\cdots$& +0.22& +0.12& +0.51& $-$0.09\\
104452&  58661& G0II        & 1.71& 2.41& 5608& 3.06&  8.4&  58.8& $-$0.30& $\cdots$& $-$0.14& $\cdots$& $\cdots$& $-$0.18\\
111812&  62763& G0III       & 1.89& 2.69& 5554& 2.91&  7.8&  64.6& $-$0.39&  2.64& $-$0.24& $\cdots$& $\cdots$& $-$0.30\\
111892&  62819& F8          & 1.36& 1.89& 5903& 3.40&  1.2&  36.3& +0.08&  1.51& +0.12& $-$0.12& +0.51& $-$0.09\\
117566&  65595& G2.5IIIb    & 1.57& 2.34& 5327& 3.10&  2.2&  11.1& $-$0.19& $\cdots$& $-$0.27& $-$0.18& +0.18& $-$0.09\\
126868&  70755& G2III       & 1.15& 1.79& 5511& 3.46&  5.3&  16.5& $-$0.31&  2.49& $-$0.18& $\cdots$& $\cdots$& $-$0.29\\
128563&  71510& F8          & 1.11& 1.63& 5927& 3.59&  0.3&  27.9& +0.23&  1.93& +0.33& +0.11& +0.53& +0.17\\
133002&  72573& F9V         & 0.99& 1.58& 5599& 3.60&  2.3&   6.6& $-$0.33& $\cdots$& $-$0.56& $-$0.33& $-$0.05& $-$0.24\\
141714&  77512& G5III-IV    & 1.59& 2.38& 5284& 3.07&  4.8&   7.7& $-$0.42&  1.15& $-$0.44& $-$0.54& +0.08& $-$0.28\\
148317&  80543& G0III       & 1.08& 1.66& 5649& 3.54&  0.3&   6.0& +0.28&  2.98& +0.14& +0.17& +0.13& +0.04\\
148743&  80840& A7Ib        & 3.80& 7.91& 7581& 2.01&  7.2&  14.6& $-$0.15& $\cdots$& $-$0.48& +0.32& $-$0.17& +0.10\\
149662&  81318& F2          & 1.30& 1.80& 6192& 3.52&  4.3& 140.1& +0.30& $\cdots$& +0.38& +1.03& +0.60& $\cdots$\\
151043&  81219& F8          & 1.05& 1.61& 5713& 3.58&  3.8&  11.7& $-$0.09& $\cdots$& +0.13& +0.19& +0.40& $-$0.08\\
151070&  81933& F5III       & 1.58& 2.15& 5977& 3.25&  2.9&  21.5& $-$0.09&  1.59& $-$0.25& $\cdots$& +0.41& +0.03\\
154319&  83114& K0          & 1.30& 1.98& 5445& 3.33&  1.1&   6.3& +0.15& $\cdots$& +0.08& +0.39& +0.19& +0.06\\
158170&  85474& F5IV        & 1.43& 1.94& 6069& 3.39&  2.4&   8.4& +0.02& $\cdots$& +0.10& +0.56& +0.46& +0.13\\
159026&  85688& F6III       & 2.31& 3.25& 6174& 2.75&  4.8& 155.1& $-$0.03& $\cdots$& +0.06& +1.20& +0.31& $-$0.38\\
160365&  86373& F6III       & 1.55& 2.09& 6083& 3.30&  9.3&  94.2& $-$0.31&  3.00& $\cdots$& +0.06& $-$0.32& $-$0.62\\
164136&  87998& F2II        & 3.03& 4.92& 6738& 2.37&  4.5&  31.7& $-$0.64&  1.72& $-$0.47& $-$0.14& $-$0.38& $-$0.06\\
164507&  88217& G5IV        & 0.78& 1.43& 5429& 3.71&  0.2&   5.6& +0.28& $\cdots$& $-$0.03& $-$0.33& $-$0.24& +0.03\\
164613&  87728& F2.5II-III  & 2.79& 4.21& 6921& 2.59&  4.3&  41.9& $-$0.22& $\cdots$& $-$0.44& $-$0.05& $-$0.17& +0.04\\
164668&  88267& G5          & 2.18& 3.37& 5027& 2.55&  4.8&   9.5& $-$0.43&  2.35& $-$0.15& $\cdots$& +0.37& $-$0.37\\
168608&  89968& F8II        & 3.20& 5.43& 6887& 2.28&  4.5&  18.2& +0.18& $\cdots$& $-$0.46& $-$0.17& $-$0.55& +0.36\\
172365&  91499& F8Ib-II     & 2.77& 4.32& 5972& 2.36&  5.2&  51.1& +0.09&  2.87& +0.09& +0.30& +0.05& +0.16\\
174464&  92488& F2Ib        & 3.37& 5.96& 7403& 2.28&  5.1&  20.7& $-$0.23& $\cdots$& $-$0.37& $-$0.14& $-$0.39& +0.24\\
180583&  94685& F6Ib-II     & 2.92& 4.69& 6253& 2.32&  5.3&   9.9& $-$0.16& $\cdots$& $-$0.38& $-$0.06& $-$0.30& +0.00\\
185758&  96757& G0II        & 2.49& 3.83& 5400& 2.41&  1.5&   8.4& +0.02& $\cdots$& $-$0.36& +0.20& $-$0.41& +0.16\\
188650&  97985& Fp          & 2.92& 4.80& 5645& 2.16&  8.8&   7.7& $-$0.64&  1.22& $-$0.47& $-$0.25& $-$0.30& $-$0.12\\
193370& 100122& F5Ib        & 3.65& 7.39& 6149& 1.77&  6.2&   9.3& $-$0.17& $\cdots$& $-$0.29& +0.38& $-$0.05& +0.12\\
194951& 100866& F1II        & 3.43& 6.21& 7418& 2.23&  5.2&  20.9& $-$0.04& $\cdots$& $-$0.38& +0.06& $-$0.26& +0.28\\
195295& 101076& F5II        & 3.09& 5.13& 6624& 2.29&  4.9&   9.2& $-$0.15&  1.46& $-$0.36& +0.20& $-$0.13& +0.11\\
196755& 101916& G5IV+...    & 0.89& 1.51& 5482& 3.64&  1.5&   6.8& $-$0.06&  1.39& $-$0.04& $-$0.31& +0.00& $-$0.15\\
198726& 102949& F5Ib        & 2.90& 4.68& 5974& 2.27&  4.7&  14.7& $-$0.18& $\cdots$& $-$0.10& +0.58& +0.32& $-$0.13\\
201078& 104185& F7.5Ib-IIv  & 3.17& 5.44& 6339& 2.16&  3.6&  12.7& $-$0.07&  1.37& $-$0.31& +0.25& $-$0.08& +0.18\\
203096& 105229& A5IV        & 3.25& 5.41& 8251& 2.54&  5.0&  27.2& $-$0.44& $\cdots$& +0.01& $-$0.13& $-$0.41& +0.49\\
208110& 108090& G0IIIs      & 1.83& 2.70& 5309& 2.89&  2.4&   7.1& $-$0.71& $\cdots$& $-$0.69& $\cdots$& $-$0.10& $-$0.71\\
210459& 109410& F5III       & 2.04& 2.78& 6262& 2.99&  5.5& 143.7& $-$0.13&  3.14& $-$0.31& +0.55& +0.19& $-$0.54\\
213306& 110991& G2Ibvar     & 3.39& 6.32& 5890& 1.88&  2.9&  13.0& +0.38& $\cdots$& $-$0.42& $-$0.02& $-$0.60& +0.36\\
220657& 115623& F8IV        & 1.62& 2.28& 5754& 3.17&  7.5&  77.3& $-$0.27&  2.37& +0.34& +0.12& $\cdots$& $-$0.50\\
\hline
\end{tabular}
\end{center}
\footnotesize
Note.\\
(1) HD number. (2) Hipparcos number. (3) Spectral type (taken from Hipparcos catalogue).
(4) Logarithm of bolometric luminosity (in unit of $L_{\odot}$).
(5) Stellar mass (in unit of $M_{\odot}$). (6) Effective temperature (in K).
(7) Logarithm of surface gravity ($\log g$ in dex,
where $g$ is in unit of cm~s$^{-2}$). (8) Microturbulent velocity (in km~s$^{-1}$).  
(9) Projected rotational velocity derived from 6145--6166~\AA\ fitting (in km~s$^{-1}$). 
(10) Differential abundance of Fe (in dex)
relative to Procyon (derived from 6145--6166~\AA\ fitting).
(11) Logarithmic number abundance of Li (in dex) expressed in the usual normalization of $A_{\rm H} = 12$.
(12)--(15) Differential abundances of C, N, O, and Na (in dex) relative to Procyon.
\end{table*}

\setcounter{table}{1}
\begin{table*}[h]
\caption{Outline of spectrum-fitting analysis in this study.}
\begin{center}
\begin{tabular}{ccccc}\hline\hline
Purpose & fitting range (\AA) & abundances varied$^{*}$ & atomic data source & figure \\
\hline
Li abundance from Li~{\sc i} 6708  & 6702--6714 &  Li, Fe & SLN98+KB95m0 & Fig.~3 \\
C abundance from C~{\sc i} 5380   & 5370--5390 & C, Ti, Fe & KB95m1 & Fig.~4 \\
N abundance from N~{\sc i} 7468  & 7457--7472 & N, Fe & KB95m2 & Fig.~5 \\
O/Na abundances from O~{\sc i} 6156--8/Na~{\sc i} 6161  & 6145--6166 & O, Na, Si, Ca, Fe & KB95 & Fig.~6 \\
\hline
\end{tabular}
\end{center}
$^{*}$ The abundances of all other elements than these were fixed in the fitting. \\
SLN98+KB95m0 --- The line list of Smith, Lambert, \& Nissen (1998) was invoked  
in the neighborhood of the Li~{\sc i} 6708 line region.
Otherwise, the data of Kurucz \& Bell (1995) were used, except that $\log gf$ values
were adjusted for Fe~{\sc i} 6703.568 ($-3.02$) as well as Fe~{\sc i} 6705.101 ($-1.02$), 
and the contributions of Ni~{\sc i} 6711.575 and Fe~{\sc i} 6712.676 were neglected.\\
KB95m1 --- All the atomic line data presented in Kurucz \& Bell (1995) were used, 
except that the contribution of Fe~{\sc i} 5382.474 was neglected.\\
KB95m2 --- All the atomic line data were taken from Kurucz \& Bell (1995), except that 
the contribution of S~{\sc i} 7468.588 was neglected.\\
KB95 --- All the atomic line data given in Kurucz \& Bell (1995) were used unchanged.
\end{table*}

\setcounter{table}{2}
\begin{table*}[h]
\caption{Adopted atomic data of the relevant lines.}
\begin{center}
\begin{tabular}{ccccccccc}\hline\hline
Line & Multiplet & Equivalent & $\lambda$ & $\chi_{\rm low}$ & $\log gf$ & Gammar & Gammas & Gammaw\\
     &  No.           &  Width  & (\AA) & (eV) & (dex) & (dex) & (dex) & (dex) \\  
\hline
Li~{\sc i} 6708 & (1) & $W_{5380}$ & 6707.756 & 0.000 & $-0.427$ & (7.69) & ($-6.54$) & ($-7.72$)\\
(6 components)&  &  & 6707.768 & 0.000 & $-0.206$ & (7.69) & ($-6.54$) & ($-7.72$)\\
         &         &          & 6707.907 & 0.000 & $-0.932$ & (7.69) & ($-6.54$) & ($-7.72$)\\
         &         &          & 6707.908 & 0.000 & $-1.161$ & (7.69) & ($-6.54$) & ($-7.72$)\\
         &         &          & 6707.919 & 0.000 & $-0.712$ & (7.69) & ($-6.54$) & ($-7.72$)\\
         &         &          & 6707.920 & 0.000 & $-0.932$ & (7.69) & ($-6.54$) & ($-7.72$)\\
\hline
C~{\sc i} 5380 & (11)  & $W_{5380}$ &  5380.337 & 7.685 & $-1.842$ & (7.89) & $-$4.66 & ($-$7.36)\\
\hline
N~{\sc i} 7468 & (3) & $W_{7468}$ & 7468.312 & 10.336 & $-0.270$  & 8.64 & $-$5.40 & ($-$7.60)\\
\hline
O~{\sc i} 6156--8 & (10) & $W_{6156-8}$ & 6155.961 & 10.740 & $-1.401$ & 7.60 & $-$3.96 & ($-$7.23)\\
(9 components)&  &  & 6155.971 & 10.740 & $-1.051$ & 7.61 & $-$3.96 & ($-$7.23)\\
         &         &          & 6155.989 & 10.740 & $-1.161$ & 7.61 & $-$3.96 & ($-$7.23)\\
         &         &          & 6156.737 & 10.740 & $-1.521$ & 7.61 & $-$3.96 & ($-$7.23)\\
         &         &          & 6156.755 & 10.740 & $-0.931$ & 7.61 & $-$3.96 & ($-$7.23)\\
         &         &          & 6156.778 & 10.740 & $-0.731$ & 7.62 & $-$3.96 & ($-$7.23)\\
         &         &          & 6158.149 & 10.741 & $-1.891$ & 7.62 & $-$3.96 & ($-$7.23)\\
         &         &          & 6158.172 & 10.741 & $-1.031$ & 7.62 & $-$3.96 & ($-$7.23)\\
         &         &          & 6158.187 & 10.741 & $-0.441$ & 7.61 & $-$3.96 & ($-$7.23)\\ 
\hline
Na~{\sc i} 6161  & (5) & $W_{6161}$ & 6160.747 & 2.104 & $-1.260$ & 7.85 & $-4.39$ & ($-7.29$)\\
\hline
\end{tabular}
\end{center}
Following columns 4--6 (laboratory wavelength, lower excitation potential, 
and $gf$ value), three kinds of damping parameters are presented in columns 7--9:  
Gammar is the radiation damping width (s$^{-1}$) [$\log\gamma_{\rm rad}$], 
Gammas is the Stark damping width (s$^{-1}$) per electron density (cm$^{-3}$) 
at $10^{4}$ K [$\log(\gamma_{\rm e}/N_{\rm e})$], and
Gammaw is the van der Waals damping width (s$^{-1}$) per hydrogen density 
(cm$^{-3}$) at $10^{4}$ K [$\log(\gamma_{\rm w}/N_{\rm H})$]. \\
All the damping parameters were taken from Kurucz \& Bell (1995), 
except for the parenthesized ones (unavailable in their compilation), 
for which the default values computed by the WIDTH9 program were assigned.
\end{table*}

\clearpage

\setcounter{figure}{0}
\begin{figure*}[p]
\begin{center}
  \includegraphics[width=10cm]{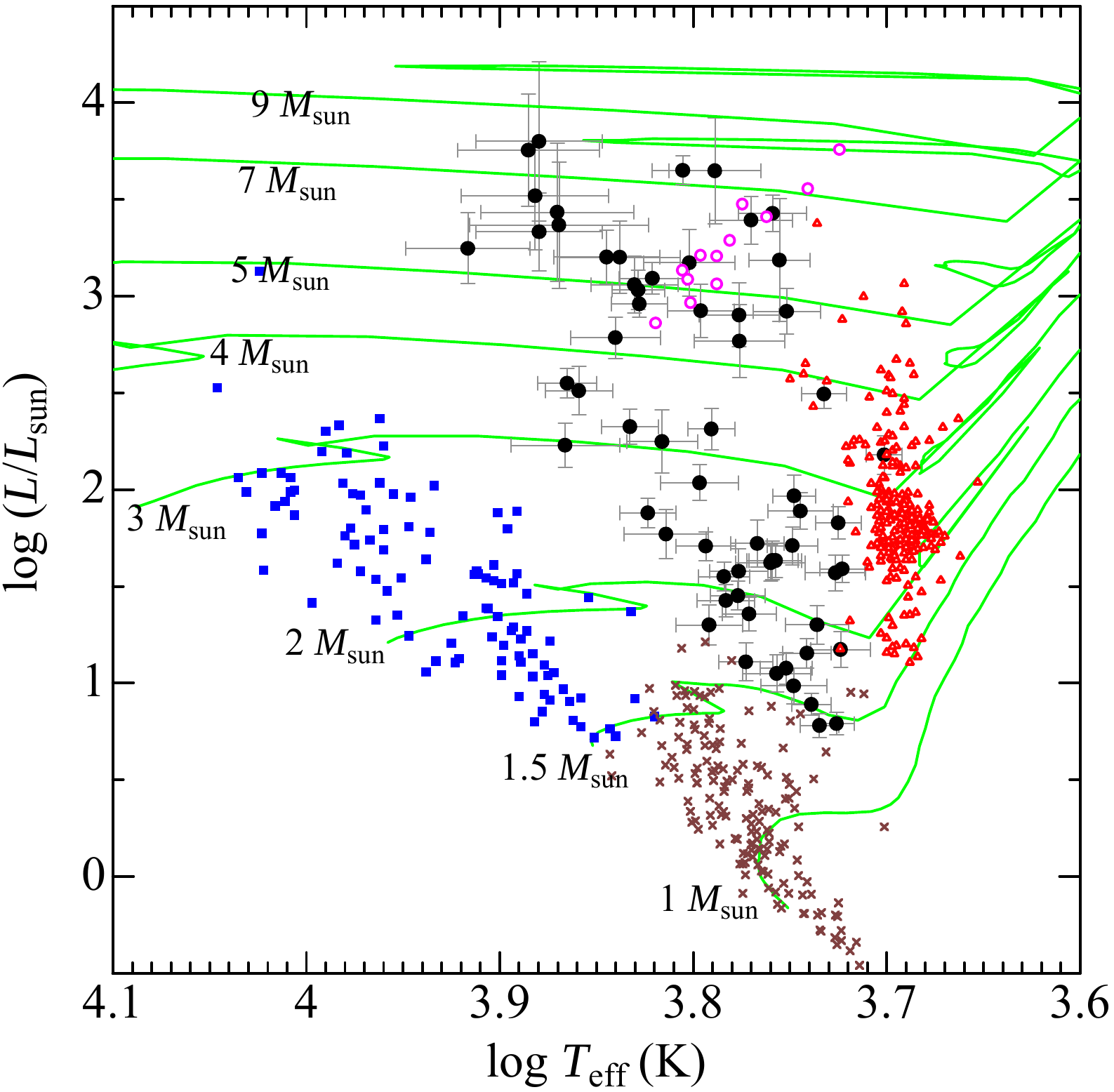}
\end{center}
\caption{
Our 62 program stars are plotted on the $\log (L/L_{\odot})$ vs. $\log T_{\rm eff}$
diagram by filled circles with error bars (see the caption of Fig.~1 in Paper~I 
for their derivation). In addition, the targets studied in Takeda \& Honda (2005)
(160 F--G--K dwarfs), Takeda et al. (2013) (12 F--G Cepheids), Takeda et al. (2015) 
(239 G--K giants), and Paper~II (100 A-type dwarfs) are also shown for comparison
by crosses, open circles, open triangles, and filled squares, respectively.
Depicted in solid lines are the theoretical solar-metallicity tracks computed by 
Lejeune and Schaerer (2001) for 8 different masses  (1, 1.5, 2, 3, 4, 5, 7, 
and 9~$M_{\odot}$). 
}
\label{fig:1}
\end{figure*}

\setcounter{figure}{1}
\begin{figure*}[p]
\begin{center}
  \includegraphics[width=12cm]{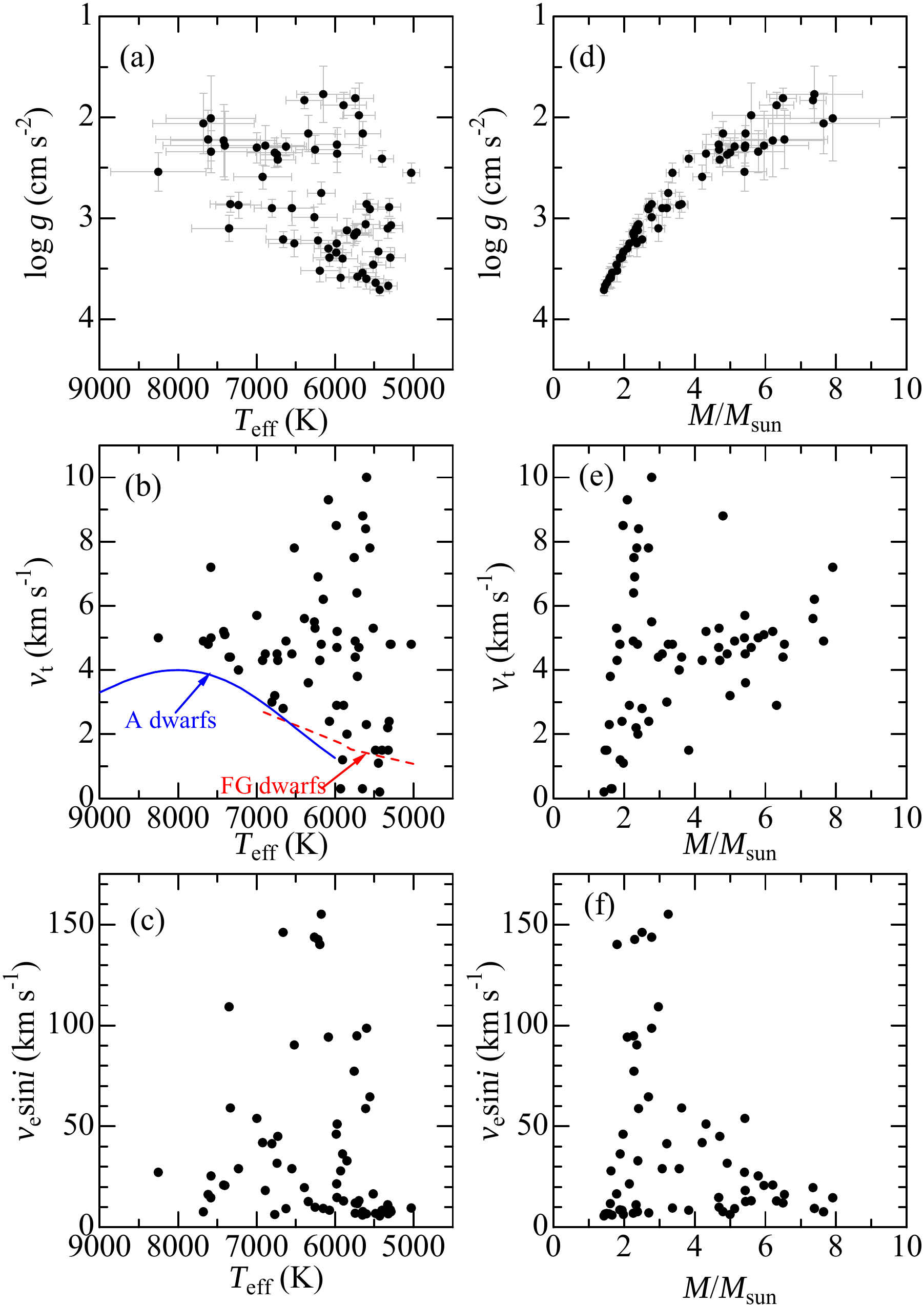}
\end{center}
\caption{Graphical display of how $\log g$ (top panels), $v_{\rm t}$ (middle panels), 
and $v_{\rm e}\sin i$ (bottom panels) correlate with $T_{\rm eff}$ (left panels) or 
$M$ (right panels). In panel (b) are also depicted the mean $v_{\rm t}$ vs. 
$T_{\rm eff}$ relations for A dwarfs and F--G dwarfs by solid and dashed lines, 
respectively (see the caption of Fig.~10 in Paper~I).
}
\label{fig:2}
\end{figure*}

\setcounter{figure}{2}
\begin{figure*}[p]
\begin{center}
  \includegraphics[width=14cm]{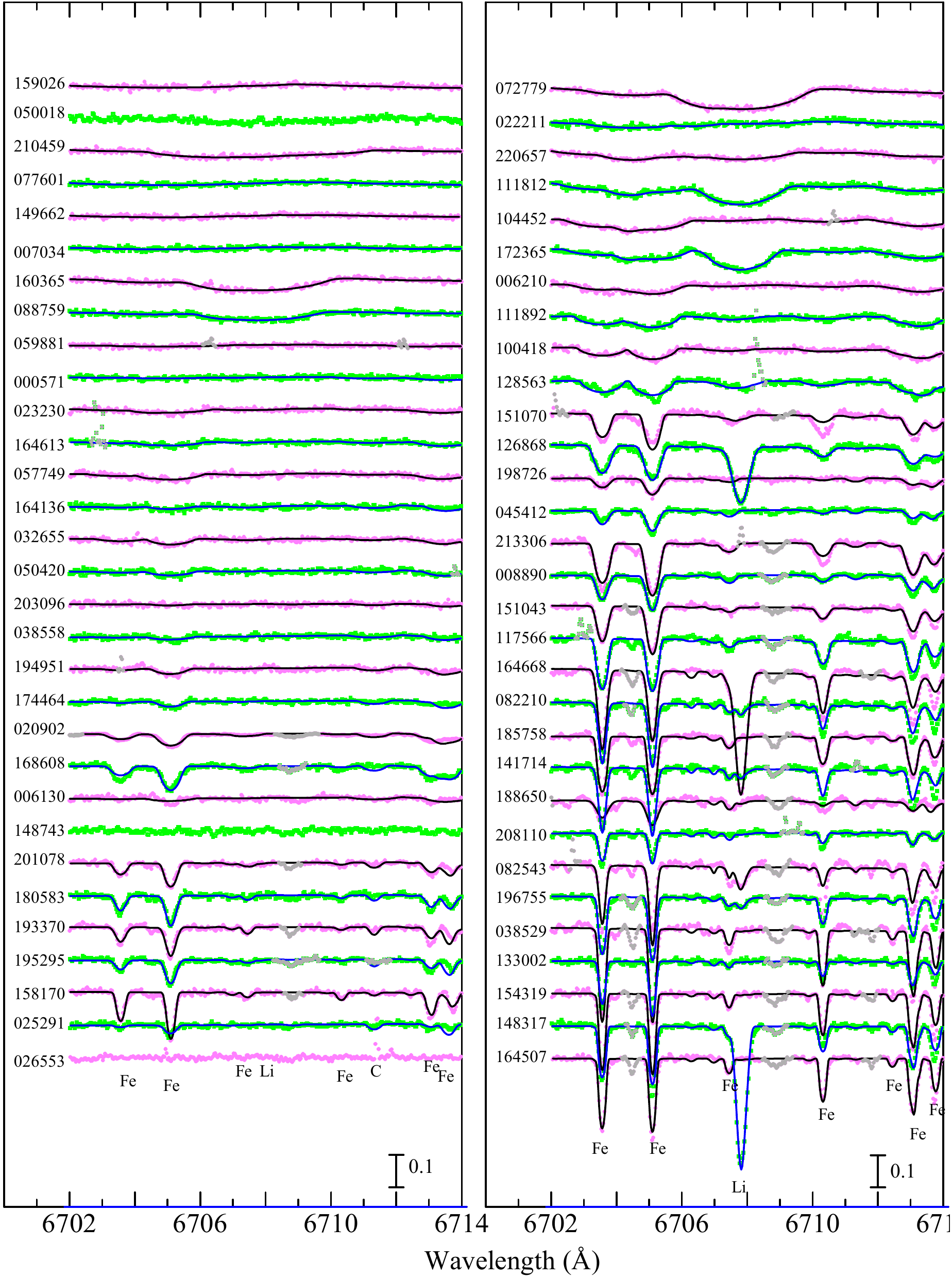}
\end{center}
\caption{
Synthetic spectrum fitting in the 6702--6714~\AA\ region 
comprising the Li~{\sc i}~6708 line. 
Shown in the left and right panels are stars of $T_{\rm eff} > 6000$~K
and those for  $T_{\rm eff} < 6000$~K, respectively.
The best-fit theoretical spectra are shown by solid lines. 
The observed data are plotted by symbols, where those used 
in the fitting are colored in pink or green, while those rejected in the 
fitting (e.g., unidentified lines or spectrum defect) are depicted in gray.
In both panels, the spectra (indicated by the HD number)
are arranged in the descending order of $v_{\rm e} \sin i$ with 
a vertical offset of 0.1 (in unit of the continuum level) 
applied to each spectrum relative to the adjacent one.  
Note that the fitting was impossible for HD~148753 and HD~026553,
for which the spectra are almost featureless. 
}
\label{fig:3}
\end{figure*}

\setcounter{figure}{3}
\begin{figure*}[p]
\begin{center}
  \includegraphics[width=14cm]{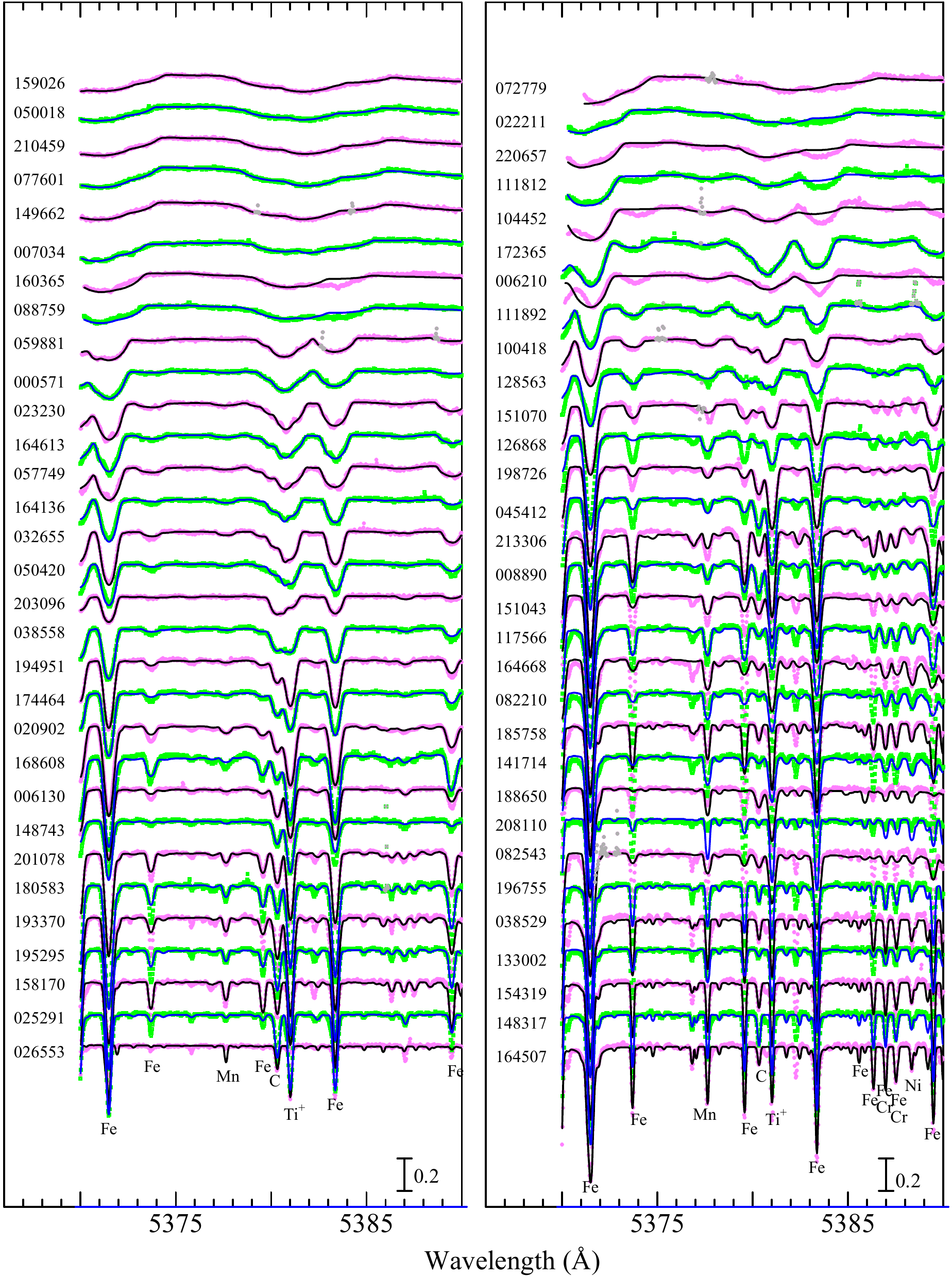}
\end{center}
\caption{
Synthetic spectrum fitting in the 5370--5390~\AA\ region 
comprising the C~{\sc i}~5380 line. 
A vertical offset of 0.2 is applied to each spectrum 
relative to the adjacent one. Otherwise, the same as in Fig.~3.
}
\label{fig:4}
\end{figure*}

\setcounter{figure}{4}
\begin{figure*}[p]
\begin{center}
  \includegraphics[width=14cm]{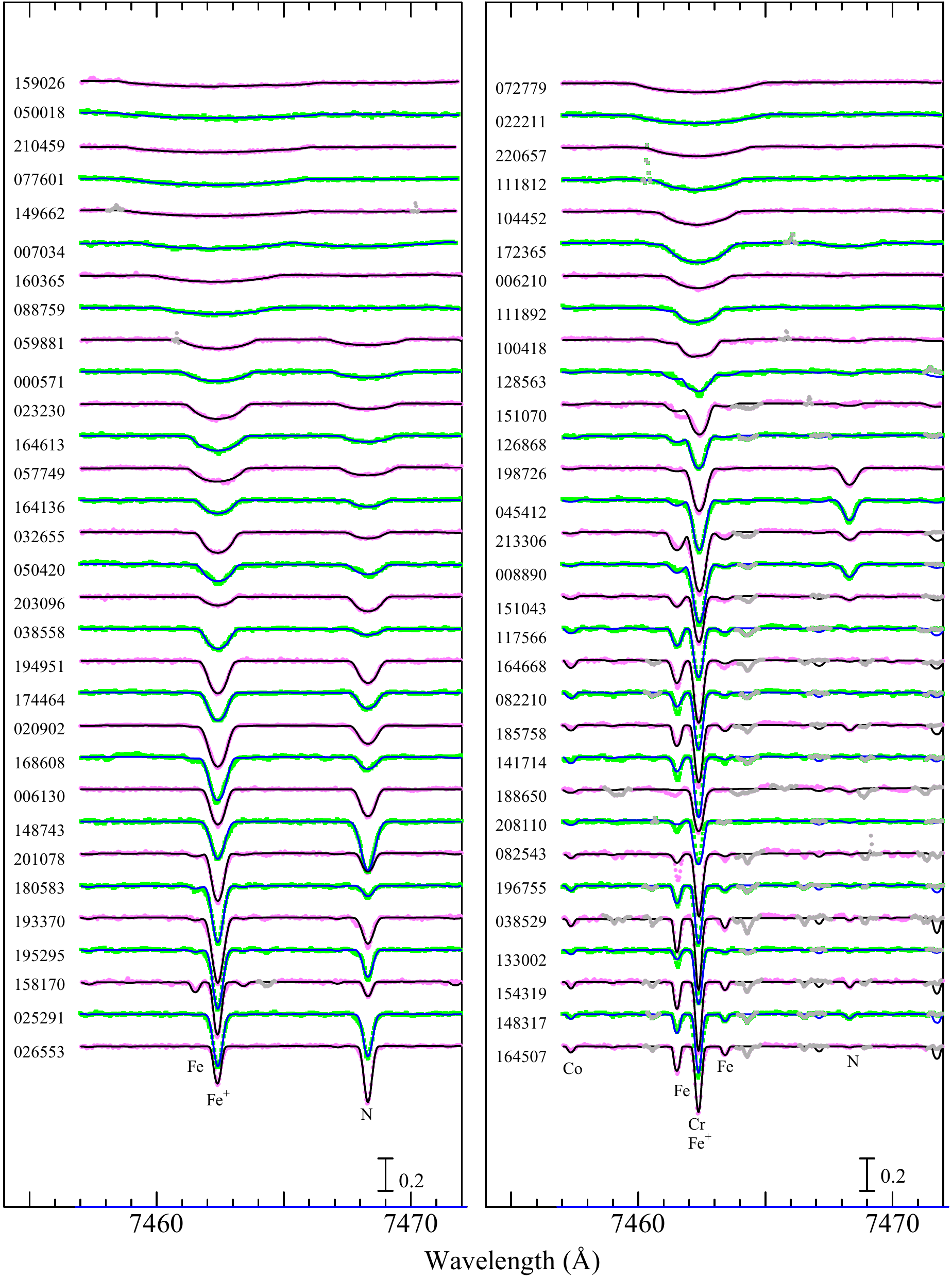}
\end{center}
\caption{
Synthetic spectrum fitting in the 7457--7472~\AA\ region 
comprising the N~{\sc i}~7468 line.  A vertical offset of 0.2 
is applied to each spectrum relative to the adjacent one. 
Otherwise, the same as in Fig.~3.
}
\label{fig:5}
\end{figure*}

\setcounter{figure}{5}
\begin{figure*}[p]
\begin{center}
  \includegraphics[width=14cm]{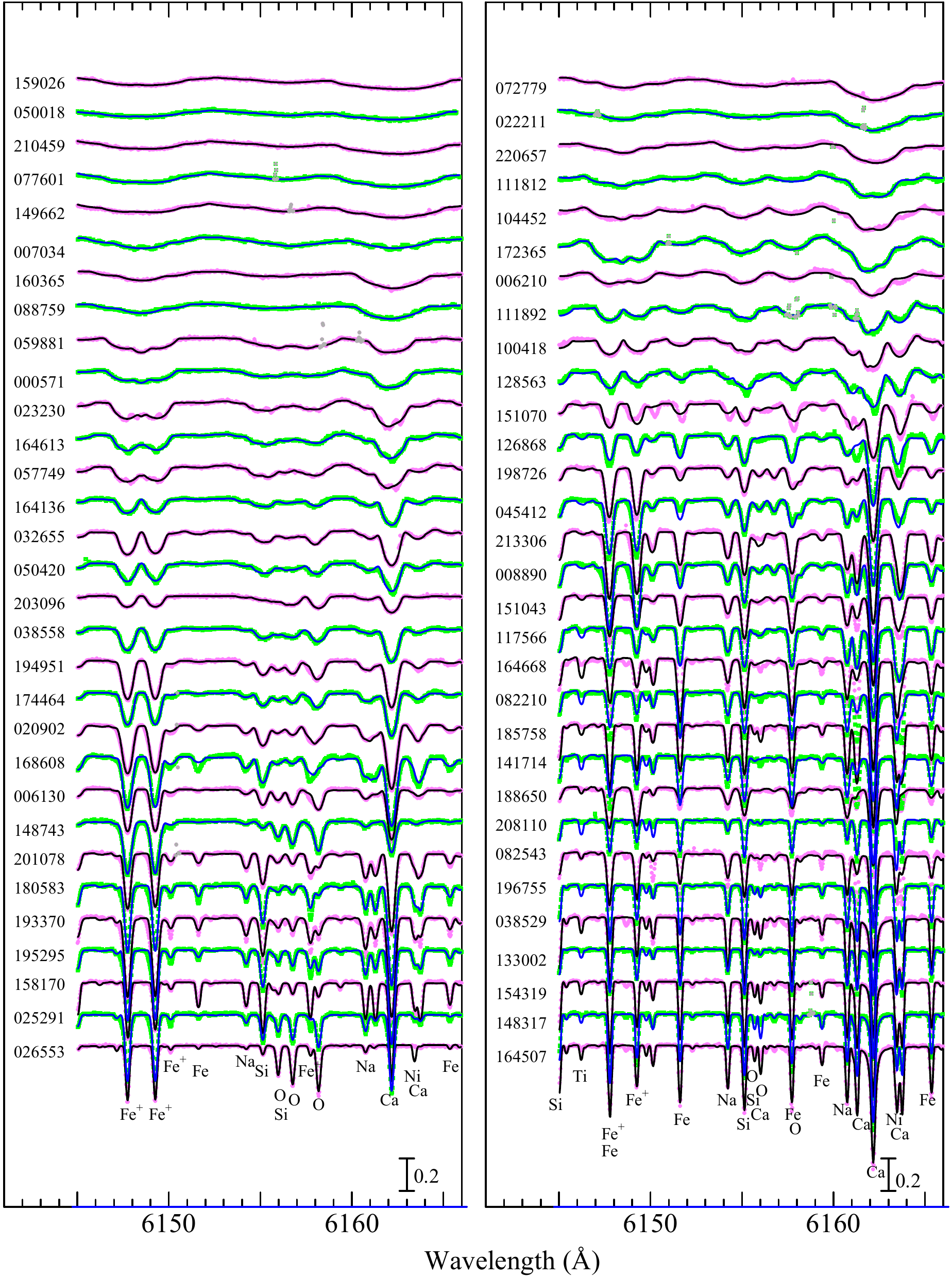}
\end{center}
\caption{
Synthetic spectrum fitting in the 6145--6166~\AA\ region 
comprising the O~{\sc i}~6156--8 and Na~{\sc i} 6161 lines. 
A vertical offset of 0.2 is applied to each spectrum relative 
to the adjacent one. Otherwise, the same as in Fig.~3.
}
\label{fig:6}
\end{figure*}

\setcounter{figure}{6}
\begin{figure*}[p]
\begin{center}
  \includegraphics[width=10cm]{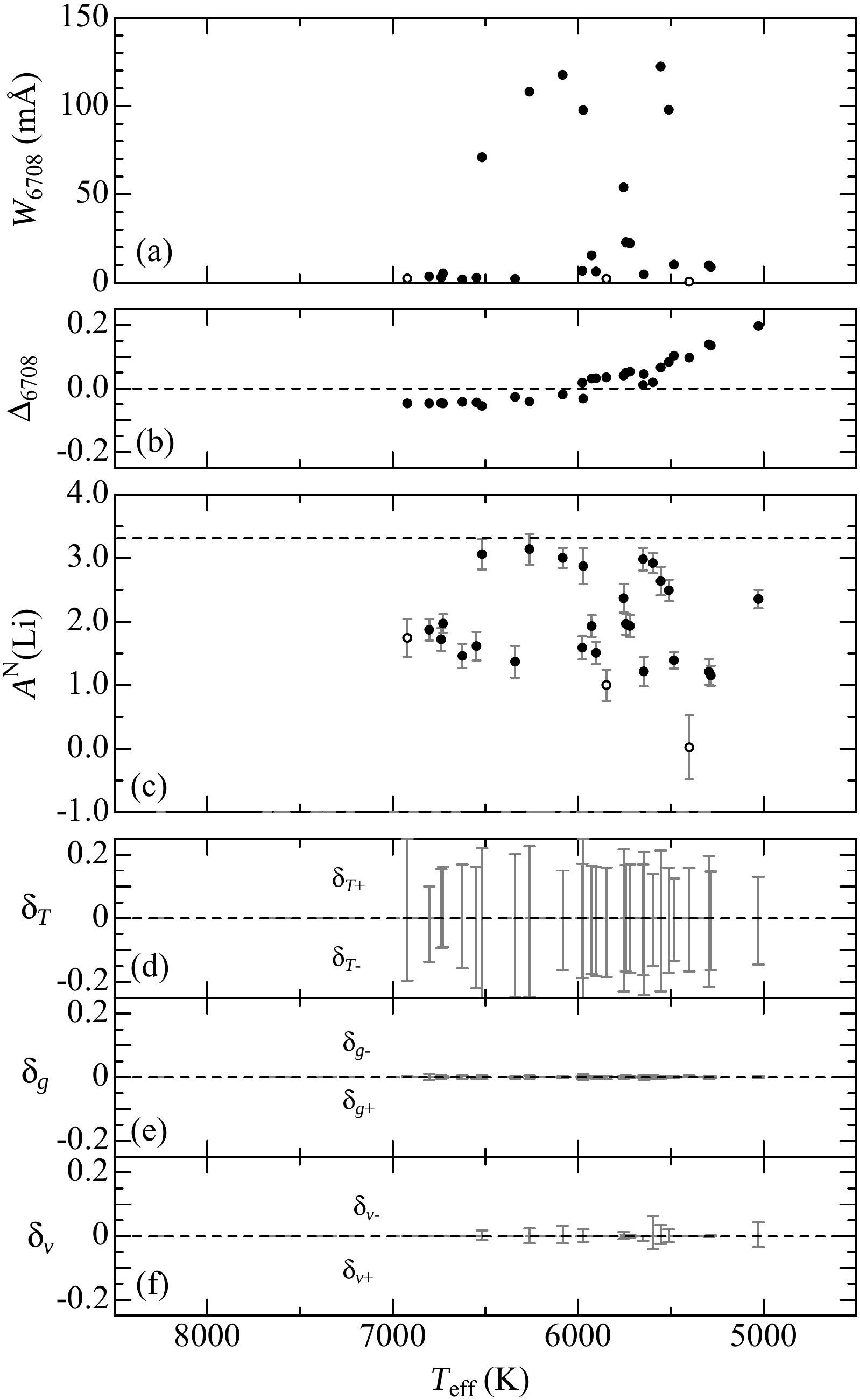}
\end{center}
\caption{
Lithium abundance and Li~{\sc i}~6708-related quantities 
plotted against $T_{\rm eff}$. 
(a) $W_{6708}$ (equivalent width of Li~{\sc i}~6708), 
(b) $\Delta_{6708}$ (non-LTE correction for Li~{\sc i}~6708),
(c) $A^{\rm N}$(Li) (non-LTE abundance derived from Li~{\sc i}~6708)
where the error bar denotes $\delta_{TgvW}$ (cf. Sect.~3.3),
(d) $\delta_{T+}$ and $\delta_{T-}$ (abundance variations 
in response to the error in $T_{\rm eff}$), 
(e) $\delta_{g+}$ and $\delta_{g-}$ (abundance variations 
in response to the error in $\log g$), 
and (f) $\delta_{v+}$ and $\delta_{v-}$ (abundance 
variations in response to perturbing the $v_{\rm t}$ value by 
$\pm$max[$0.3 v_{\rm t}$, 1~km~s$^{-1}$]).
The data shown in open circles in panels (a) and (c) denote unreliable
results which were eventually discarded (cf. Sect.~3.3).
The reference solar-system Li abundance of $A_{\rm ss}$ = 3.31 is indicated 
by the horizontal dashed line in panel (c). 
}
\label{fig:7}
\end{figure*}

\setcounter{figure}{7}
\begin{figure*}[p]
\begin{center}
  \includegraphics[width=10cm]{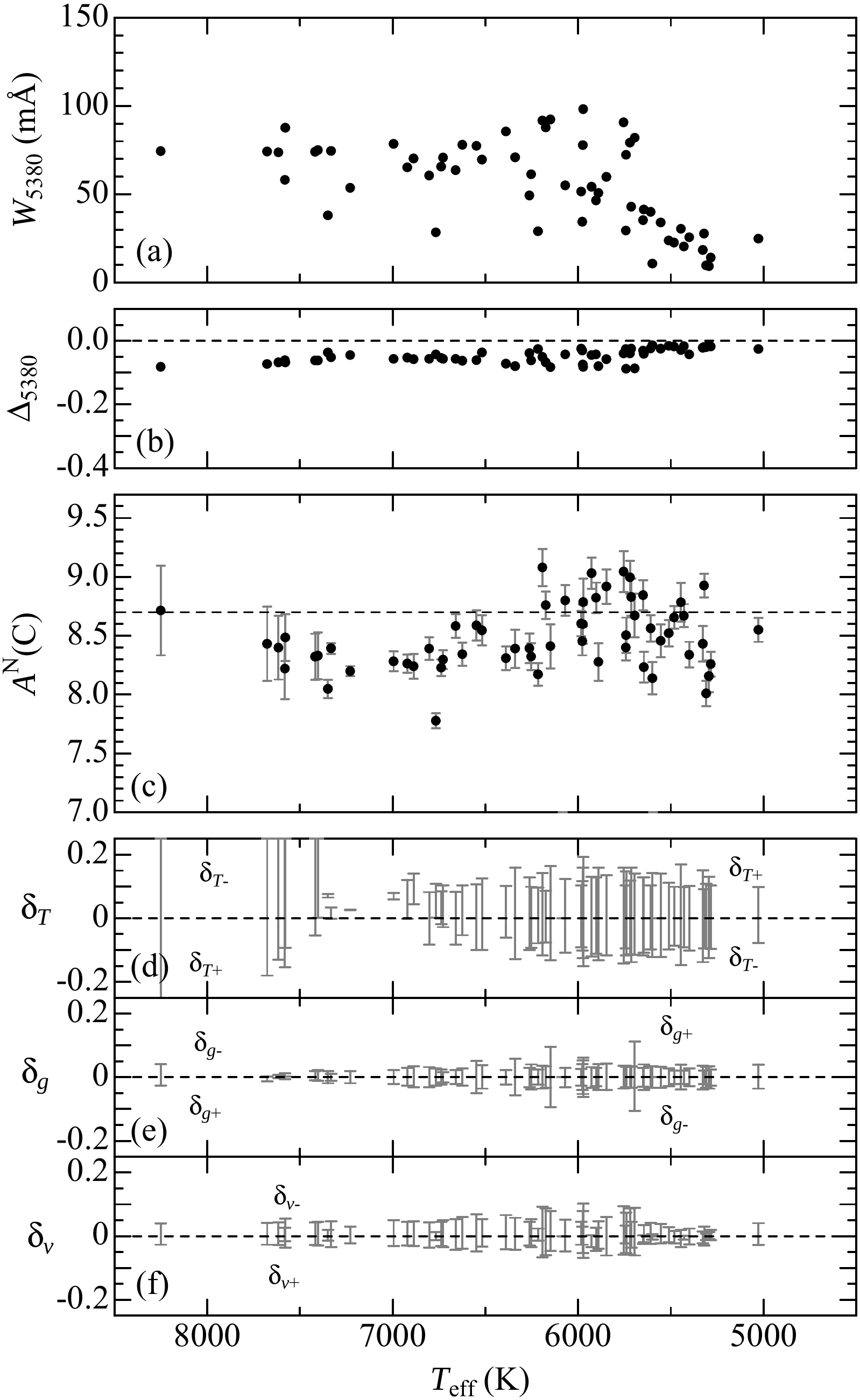}
\end{center}
\caption{
Carbon abundance and C~{\sc i}~5380-related quantities 
plotted against $T_{\rm eff}$. 
The reference C abundance of $A^{\rm N}$(C:Procyon) = 8.70 is 
indicated by the horizontal dashed line in panel (c). 
Otherwise, the same as in Fig.~7. 
}
\label{fig:8}
\end{figure*}

\setcounter{figure}{8}
\begin{figure*}[p]
\begin{center}
  \includegraphics[width=10cm]{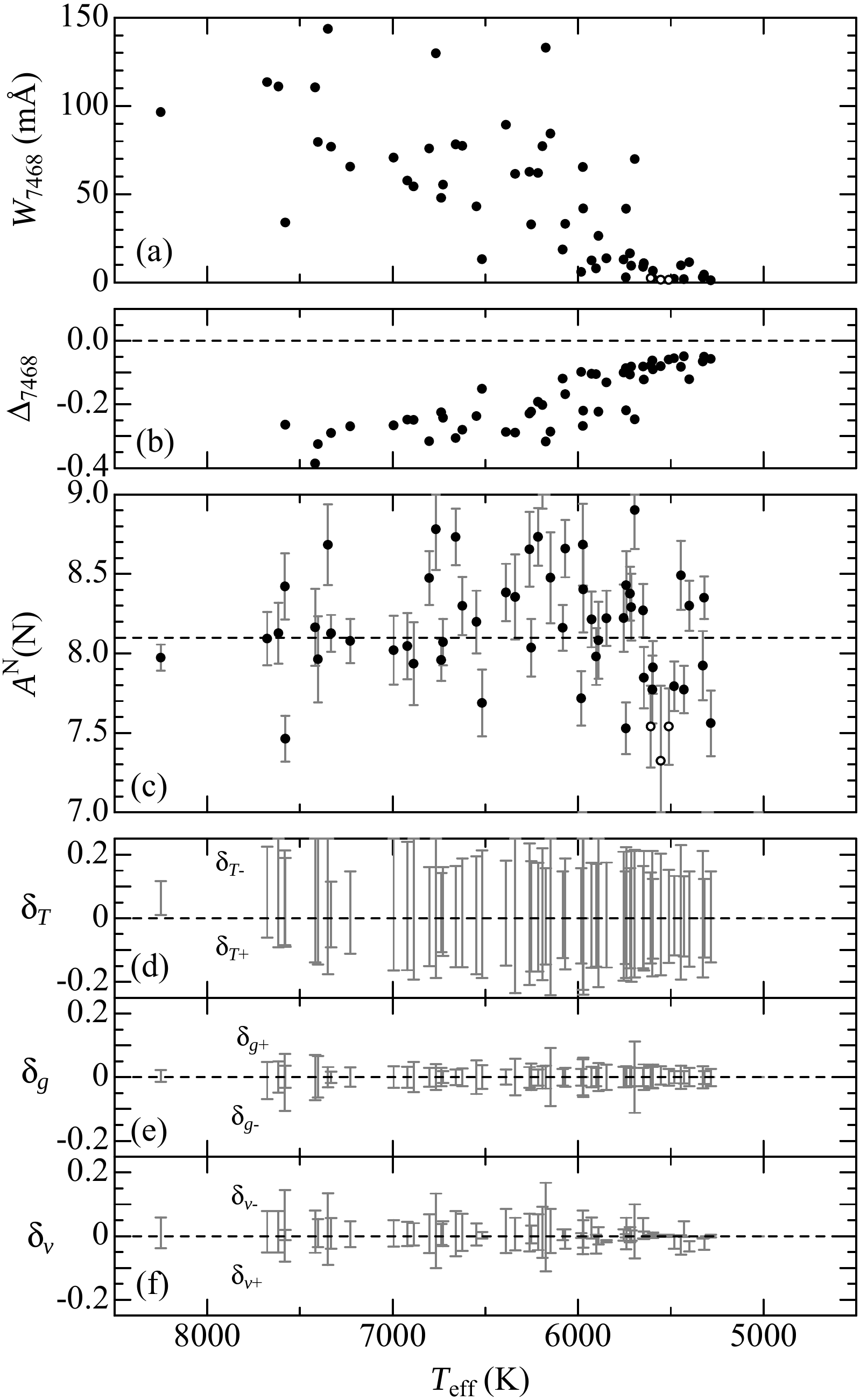}
\end{center}
\caption{
Nitrogen abundance and N~{\sc i}~7486-related quantities 
plotted against $T_{\rm eff}$. 
The reference N abundance of $A^{\rm N}$(N:Procyon) = 8.10 is 
indicated by the horizontal dashed line in panel (c). 
Otherwise, the same as in Fig.~7. 
}
\label{fig:9}
\end{figure*}

\setcounter{figure}{9}
\begin{figure*}[p]
\begin{center}
  \includegraphics[width=10cm]{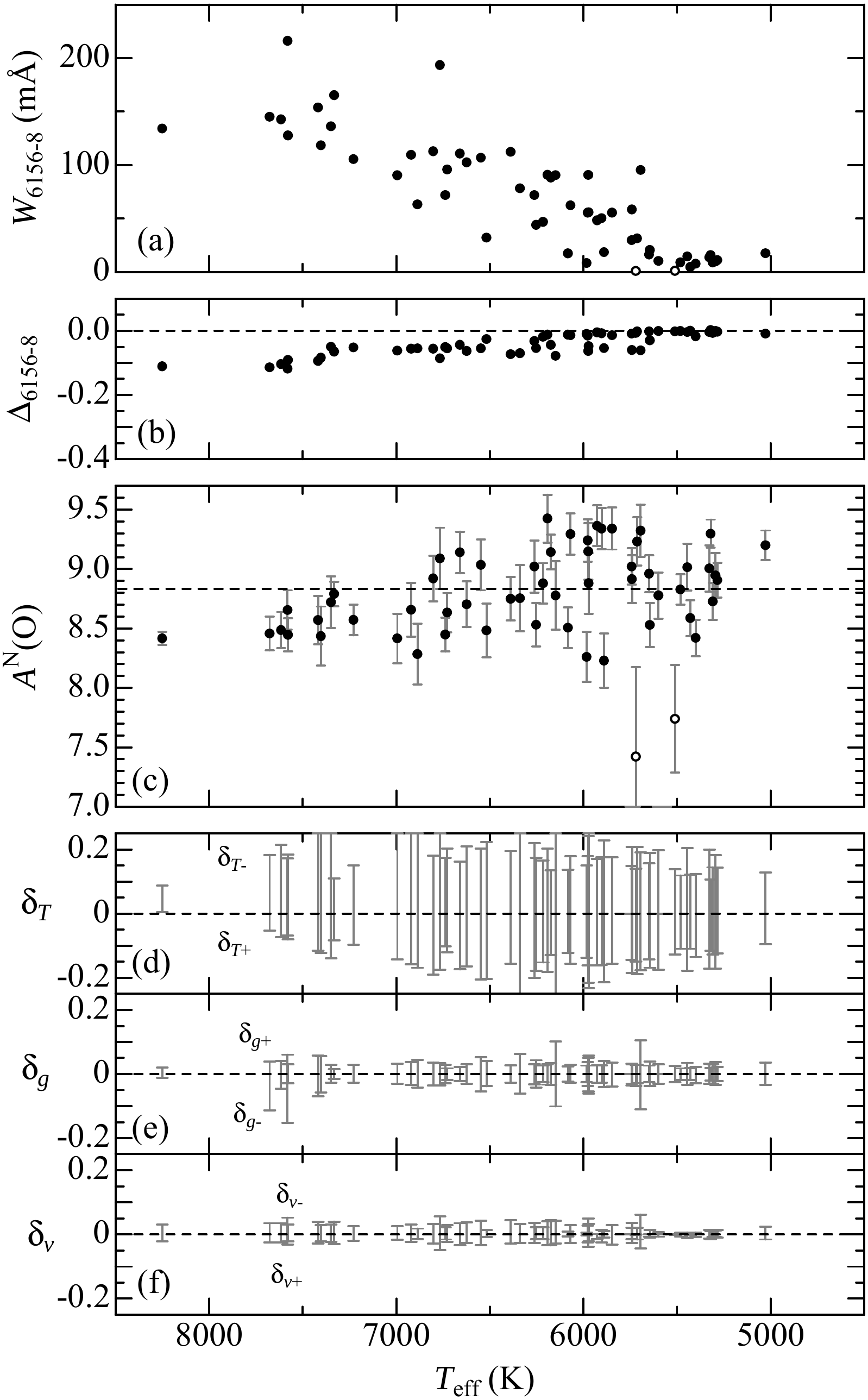}
\end{center}
\caption{
Oxygen abundance and O~{\sc i}~6156--8-related quantities 
plotted against $T_{\rm eff}$. 
The reference O abundance of $A^{\rm N}$(O:Procyon) = 8.83 is 
indicated by the horizontal dashed line in panel (c). 
Otherwise, the same as in Fig.~7. 
}
\label{fig:10}
\end{figure*}

\setcounter{figure}{10}
\begin{figure*}[p]
\begin{center}
  \includegraphics[width=10cm]{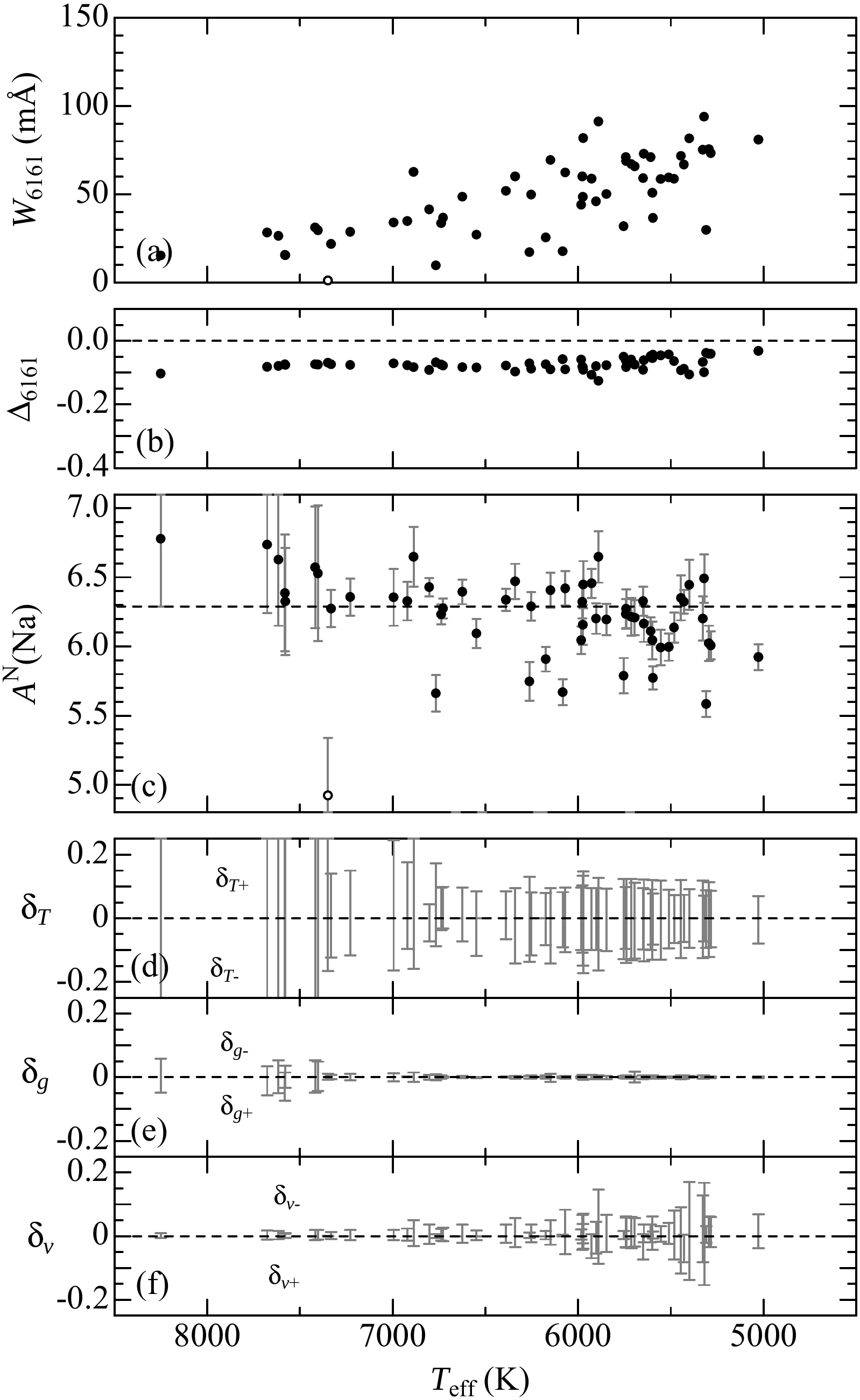}
\end{center}
\caption{
Sodium abundance and Na~{\sc i}~6161-related quantities 
plotted against $T_{\rm eff}$. 
The reference Na abundance of $A^{\rm N}$(Na:Procyon) = 6.29 is 
indicated by the horizontal dashed line in panel (c). 
Otherwise, the same as in Fig.~7. 
}
\label{fig:11}
\end{figure*}

\setcounter{figure}{11}
\begin{figure*}[p]
\begin{center}
  \includegraphics[width=14cm]{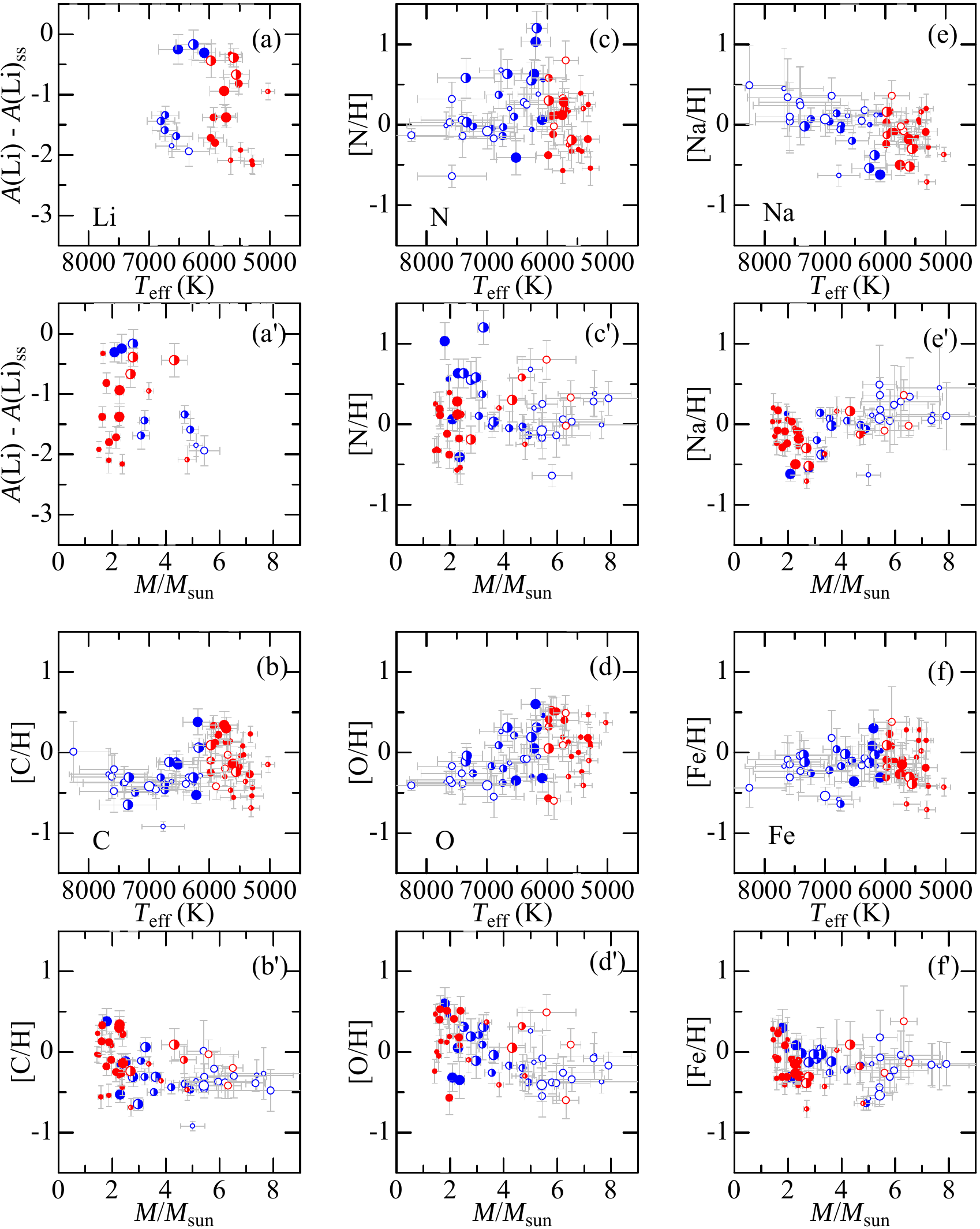}
\end{center}
\caption{
These six figure sets (a--f, each consisting of 2 panels)
illustrate how the relative abundances of six elements 
(L, C, N, O, Na, and Fe) with respect to the reference abundances 
(the solar system abundance for Li; otherwise, the abundances of Procyon) 
depend upon $T_{\rm eff}$ (upper panel) or $M$ (lower panel).
Different mass classes are discriminated by the symbol shape:
filled circles $\cdots$ $M < 2.5 M_{\odot}$, 
half-filled circles $\cdots$ $2.5 M_{\odot} < M < 5 M_{\odot}$, and
open circles $\cdots$ $5 M_{\odot} < M$.
Different $v_{\rm e}\sin i$ classes are discriminated by the symbol size:
small $\cdots$ $v_{\rm e}\sin i < 10$~km~s$^{-1}$,
medium $\cdots$ 10~km~s$^{-1} < v_{\rm e}\sin i < 50$~km~s$^{-1}$, and
large $\cdots$ 50~km~s$^{-1} < v_{\rm e}\sin i$.
Stars of higher $T_{\rm eff}$ ($>6000$~K) and lower $T_{\rm eff}$ ($< 6000$~K) 
are colored in blue and red, respectively.
Regarding the meanings of the error bars, see Sect.~2 ($\sigma$ for $T_{\rm eff}$ 
as well as $M$) and Sect.~3.3 ($\delta_{TgvW}$ for the abundances).
}
\label{fig:12}
\end{figure*}

\setcounter{figure}{12}
\begin{figure*}[p]
\begin{center}
  \includegraphics[width=14cm]{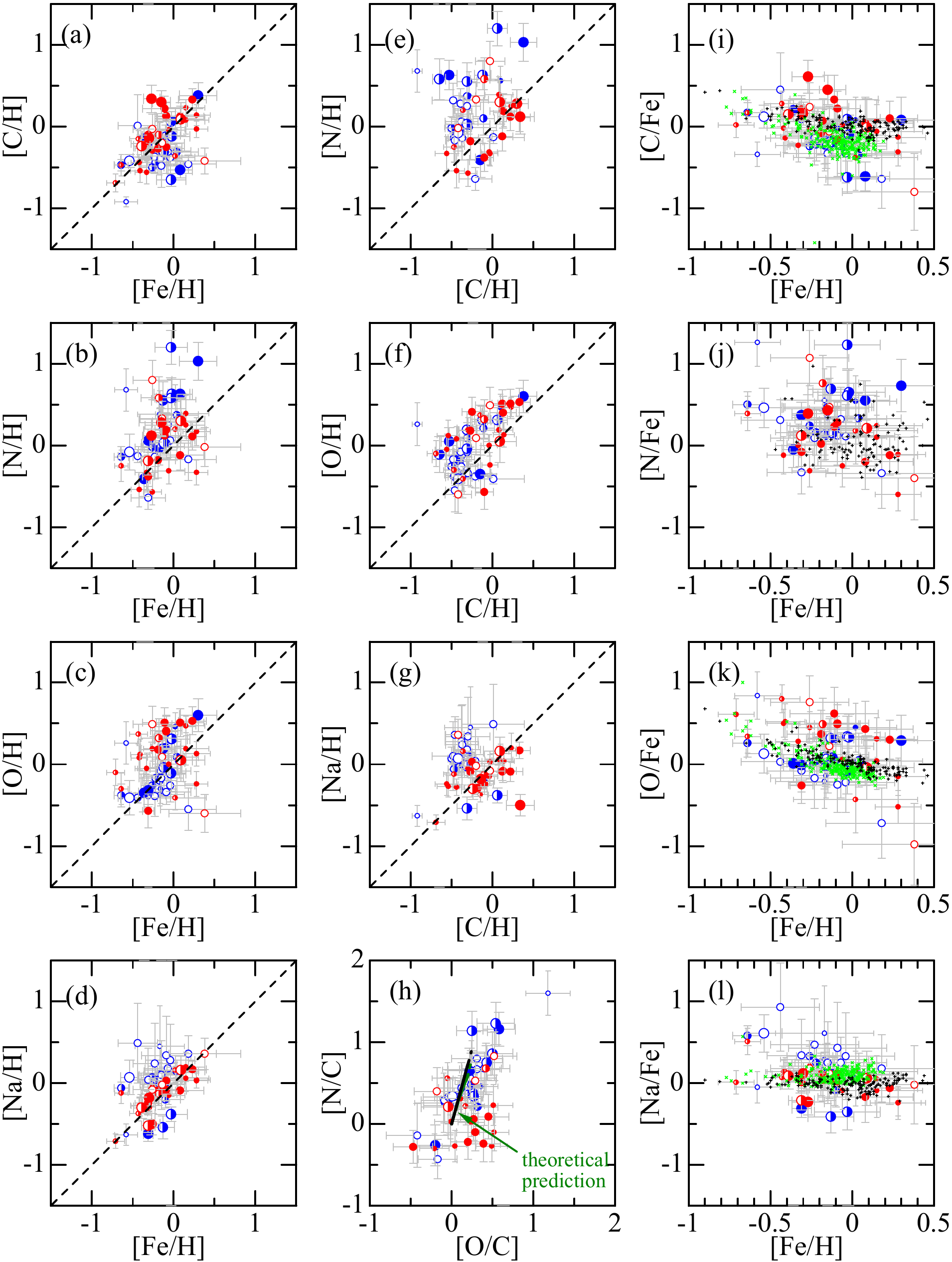}
\end{center}
\caption{
The left-hand and middle panels demonstrate the mutual correlations between 
the [C/H], [N/H], [O/H], [Na/H], and [Fe/H] values derived for our 62 program stars: 
(a) C vs. Fe, (b) N vs Fe, (c) O vs. Fe, (d) Na vs. Fe, (e) N vs. C,  (f) O vs. C, 
and (g) Na vs. C. The relationship between [N/C] and [O/C] ratios is shown in
panel (h), where the thick line is the theoretically predicted locus computed 
by Lagarde et al. (2012). [Although four different loci of solar metallicity models
corresponding to the combination of (2~$M_{\odot}$, 6~$M_{\odot}$) and (standard mixing,
non-standard mixing including thermohaline+rotational mixing) are overplotted here,
they are hardly discernible from each other.] 
In the right-hand panels (i)--(l) are plotted the abundance ratios 
([C/Fe], [N/Fe], [O/Fe], and [Na/Fe]) against [Fe/H], where the relations for 
160 FGK dwarfs and 239 GK giants (cf. Fig.~10 in Takeda et al. 2015, while 
the [N/Fe] results for FGK dwarfs were taken from Takeda \& Honda 2005) are 
also shown in black Greek crosses (+) and light-green St. Andrew's crosses ($\times$), 
respectively. See the caption of Fig.~12 for the meanings of the other symbols.
The error bars attached to the data points in panels (a)--(g) denote $\delta_{TgvW}$
described in Sect.~3.3, 
while the errors for [N/C] or [O/C] shown in panel (h) 
as well as those for [X/Fe] in panels (i)--(l)  were evaluated 
as the root-sum-square of relevant two $\delta_{TgvW}$ values.
}
\label{fig:13}
\end{figure*}

\setcounter{figure}{13}
\begin{figure*}[p]
\begin{center}
  \includegraphics[width=11cm]{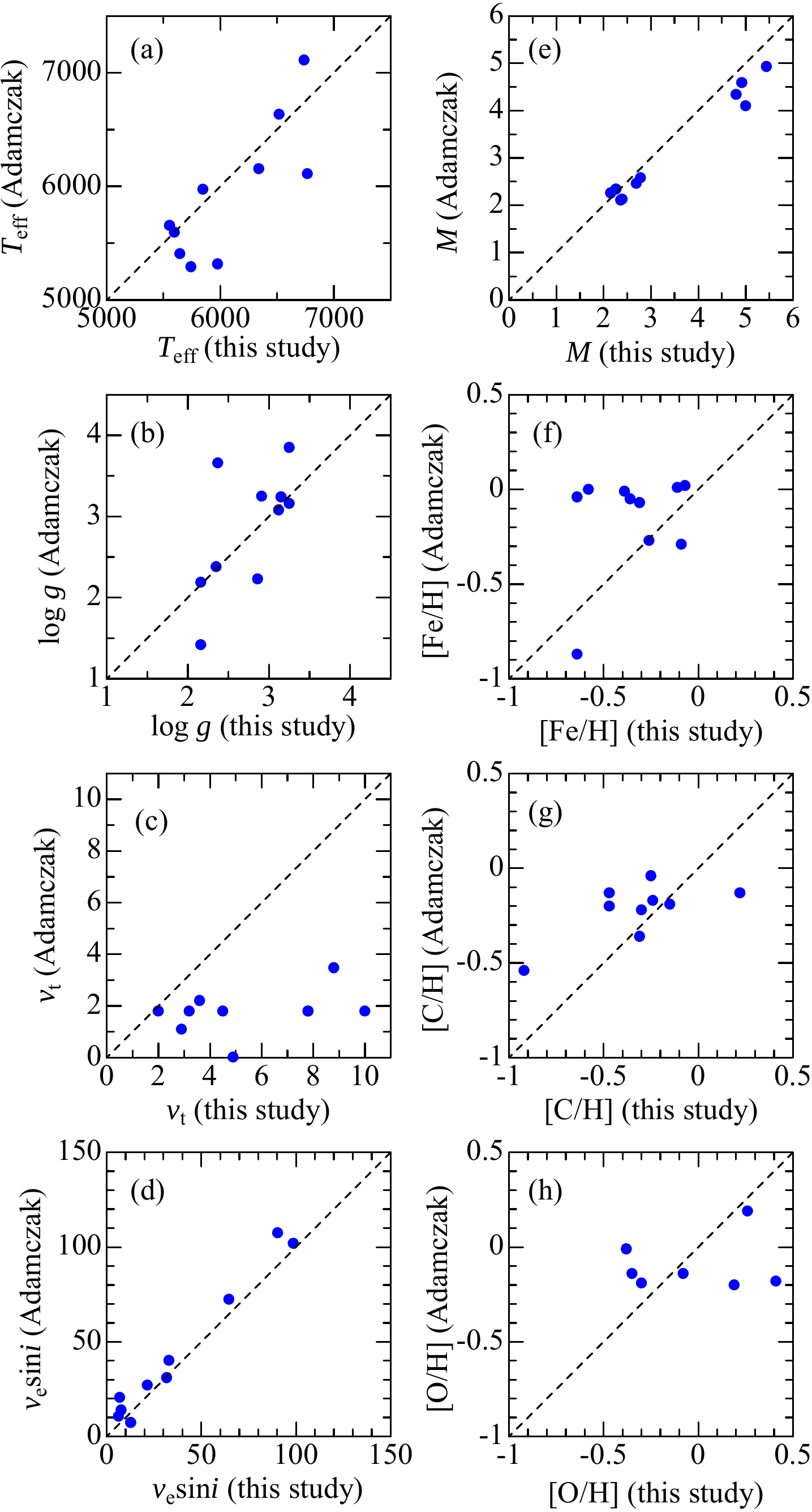}
\end{center}
\caption{
Comparison of the adopted stellar parameters and the resulting abundances in this study 
with those of Adamczak \& Lambert (2014): (a) $T_{\rm eff}$, (b) $\log g$, (c) $v_{\rm t}$,
(d) $v_{\rm e}\sin i$, (e) $M$, (f) [Fe/H], (g) [C/H], and (h) [O/H].
}
\label{fig:14}
\end{figure*}

\appendix

\section{On the source of abundance errors}

Our analysis of 62 Hertzsprung-gap stars resulted in considerable scatters 
in the [C/H], [N/H], [O/H], [Na/H] and [Fe/H] abundances. The cause of such 
apparently large dispersions may deserve some examination.
We here consider two possibilities: (i) errors in the atmospheric parameters 
and (ii) unusual atmospheric condition caused by stellar activity. 

\subsection{Uncertainties in atmospheric parameters}

As one of the touchstones to judge the reasonability of our parameters adopted 
for each star ($T_{\rm eff}$ from reddening-corrected $B-V$ color, $\log g$ from 
parallax-based/extinction-corrected luminosity along with stellar mass estimated 
from evolutionary tracks; cf. Sect.~3 in Paper~I), it is worthwhile to examine 
how they are compared with various published values determined in different ways. 
Although such a comparison was already tried by using the data taken from 
SIMBAD database in Paper~I (see Fig.~2 therein), we here make use of the 
PASTEL database compiled by Soubiran et al. (2010) for the data source of 
$T_{\rm eff}$, $\log g$, and [Fe/H], the comparisons of which are shown 
in  Fig.~A1. The mean differences ($\pm \sigma$: standard deviation) are
$\langle T_{\rm eff} \rangle$ (PASTEL$-$ours) = $26 (\pm 380)$~K,
$\langle \log g \rangle$ (PASTEL$-$ours) = $+0.01 (\pm 0.45)$~dex, and
$\langle$[Fe/H]$\rangle$ (PASTEL$-$ours) = $+0.15 (\pm 0.18)$~dex.
Meanwhile, regarding the errors we estimated for $T_{\rm eff}$,
$\log g$, and [Fe/H] (cf. Sect. 2 for $T_{\rm eff}$ and $\log g$; 
Sect. 3.3 for [Fe/H]), their averages over all 62 stars are 280~K 
(ranging from 108~K to 788~K), 0.14~dex (ranging from 0.06~dex to 0.42~dex), 
and 0.19~dex (ranging from 0.06~dex to 0.44~dex), respectively.
Comparing these with $\sigma$(PASTEL$-$ours) values mentioned above,
we see that the discrepancy in $\log g$ is appreciably larger than our
estimated uncertainties, which may reflects the intrinsic difficulty in
determining the surface gravity for such high-luminosity stars. This situation is 
markedly different from the case of dwarf stars, where the spectroscopic
$\log g$ and the luminosity-based $\log g$ based on the Hipparcos parallaxes\footnote{
We also checked how the Hipparcos parallaxes (van Leeuwen 2007) we used for
evaluation of $\log g$ are compared with the Gaia DR2 parallaxes recently
published (https://www.cosmos.esa.int/web/gaia/dr2). 
Among the 62 program stars, Gaia data are missing for 2 stars 
(HD~8890, 20902) and negative for 2 stars (HD~168608, 213306). Otherwise,
the differences are a few to several tens per cent for the remaining stars 
($ < 10$\% for more than the half of them), except that especially large 
discrepancies by more than a factor 2 are found for 4 stars (HD~32665,
45412, 148713, and 164136).} (as adopted this study) agree with each other 
(see, e.g., Fig.~5 in Takeda et al. 2005). However, we do not consider
that ambiguities in $\log g$ would lead to serious abundance errors,
because of the insignificant gravity-sensitivity as can be seen in Fig.~7e--11e.

Besides, we compared our parameters ($T_{\rm eff}$, $\log g$, $v_{\rm t}$, 
and [Fe/H]) of comparatively sharp-lined G-type stars with the values
spectroscopically determined by Takeda et al. (2005; 2 stars in common) 
and Takeda et al. (2008; 5 stars in common) based on the equivalent widths 
of Fe~{\sc i} and Fe~{\sc ii} lines, as displayed in Fig.~A2. This figure 
indicates that, while a reasonable agreement is mostly confirmed for
$T_{\rm eff}$, $\log g$, and [Fe/H], a distinct discrepancy is observed
between our adopted $v_{\rm t}$ values of HD~82210 and HD~188650 (4.8 and 
8.8~km~s$^{-1}$, which were derived from O~{\sc i} lines in Paper~I) 
and those determined by using Fe lines (1.1 and 2.2~km~s$^{-1}$; cf. Fig.~A2c).
Since such large scale $v_{\rm t}$ values have occasionally been reported 
for evolved stars above the main sequence (see, e.g., the references quoted in 
Sect.~4.2 of Gray 1978), we would reserve the decision of which is correct. 
Yet, we should keep in mind a possibility that the prominently large 
$v_{\rm t}$ values up to $\lesssim 10$~km~s$^{-1}$ around $T_{\rm eff} \sim 6000$~K 
(cf. Fig.~2b) may be spurious (i.e., erroneous overestimation), even though 
these are the unique solutions which could satisfy the non-LTE abundance 
consistency between the O~{\sc i} 7771--5 and O~{\sc i} 6156--8 features. 
For example, such an overestimation of $v_{\rm t}$ from O~{\sc i} lines may 
happen when the stronger O~{\sc i} 7771--5 triplet suffers an extra intensification 
by a chromospheric temperature rise in the upper atmosphere (cf. Takeda 1995b). 
(Interestingly, both of these HD~82210 and HD~188650 show core emissions in the core 
of the Ca~{\sc ii} K line; cf. Fig.~A3). In any event, since the resulting abundances 
are insensitive to a change in $v_{\rm t}$ (because the relevant lines 
are fairly weak and the thermal velocity is large for these light elements;
cf. Fig.~7f--11f), it is unlikely that ambiguities in $v_{\rm t}$ can 
cause significant abundance errors in the present case.

Considering what has been described above and that the most important parameter
affecting the abundance is $T_{\rm eff}$ (cf. Fig.~7d--11d), we feel that 
our error bars shown in Fig.~12 and Fig.~13 (which are determined mainly
by ambiguities in $T_{\rm eff}$ typically by $\sim 300$~K on the average) 
are not seriously misvalued in the general sense, although some cases of 
considerably wrong abundances caused by exceptionally large parameter 
errors can not be excluded.  

\subsection{Possibility of influential chromospheric activity}

In connection with the large scatter observed in our abundances (e.g., [X/Fe] vs. [Fe/H]
diagrams shown in Fig.~13i--l), it is worth noting that Takeda \& Tajitsu (2017) similarly 
reported anomalously large dispersion in the C and O abundances (derived from C~{\sc i} 5380 
and O~{\sc i} 7771--5 lines of high excitation) in a fraction of Li-rich G-type giants 
(cf. Figs.~14a and 14c therein), which they considered nothing but a superficial 
phenomenon caused by high chromospheric activity. 
Regarding such activity-related effects, two mechanisms may be possible. 
The first is the temperature rise in the upper layer, by which high-excitation 
lines of dominant-stage species (such as C~{\sc i}, N~{\sc i}, or O~{\sc i})\footnote{
Although all these C~{\sc i} (I.P.=11.3~eV), N~{\sc i} (I.P.=14.5~eV), and O~{\sc i} 
(I.P.=13.6eV) are still the dominant population stage in the line-forming region 
of F--G giants/supergiants ($T_{\rm eff} \lesssim 7000$~K), this argument does not 
hold any more for A-type stars, where C~{\sc i} turns into minor population
species, though N~{\sc i} and O~{\sc i} still remain as dominant population.}  
can be strengthened (see Sect.~9 in Takeda \& Tajitsu 2017).
The second is the enhanced UV radiation, by which the lines of minor-population
species (such as Fe~{\sc i} or Na~{\sc i}) are weakened due to the overionization
effect. The apparent underabundances of heavier elements (such as Fe) in 
the secondary component of Capella (G0~{\sc iii} giant of high activity) compared 
to the primary (normal G8~{\sc iii} giant) concluded by Takeda et al. (2018b) 
can be attributed to this mechanism (cf. Sect.~6.3 therein). 
 
Meanwhile, as to the abundances obtained in this study, we see from Fig.~12 that 
(1) apparently deviated data points are prominently large [C/H], [N/H], and [O/H] 
(even up to $\sim +1$) and unusually small [Na/H] (down to $\sim -0.7$), 
(2) the large abundance dispersion tends to be seen in lower $T_{\rm eff}$ ($< 6500$~K) stars\footnote{
Note that a similar argument is possible for $M$ (i.e., large scatter tends to be
observed at lower $M$ of $\sim$~2--3~$M_{\odot}$). This is because $T_{\rm eff}$ and $M$ 
are closely correlated with each other, since lower $T_{\rm eff}$ stars tend to 
be of higher $\log g$ (cf. Fig.~2a) and thus of lower $M$ (see the relation 
in Fig.~2d).} cooler than the granulation boundary (see, e.g., Gray 2005),
and (3) stars showing large deviation tend to have comparatively large $v_{\rm e}\sin i$.
Therefore, it may be a ponderable interpretation that high chromospheric activity caused 
by rapid rotation in cooler stars having deep convection zone (in which rotation-induced 
dynamo mechanism can operate) would have especially intensified the C~{\sc i} 5380, 
N~{\sc i}~7460, and O~{\sc i}~6156--8 lines (all high-excitation lines of dominant 
species) and weakened the Na~{\sc i} 6161 line (minor population species).
 
Motivated by this idea, we checked the Ca~{\sc ii} K line at 3933.663~\AA\ for each 
star in order to search for any core emission indicative of strong chromospheric
activity, making use of the fact that our spectral data cover wide wavelength range
down to the violet region. The spectra in the 3920--3950~\AA\ portion are displayed 
in Fig.~A3 for all 62 stars. Although the S/N ratios at the core of this strong 
line are insufficient in not a few cases (because of the decreased detector sensitivity 
in this short wavelength region along with the considerably low flux level), 
which makes examination of weak feature difficult, the following characteristics 
can be read from this figure. 
\begin{itemize}
\item
Clear core emission is hardly seen for stars of higher $T_{\rm eff}$ ($> 6000$~K) 
(cf. the left panel of Fig~A3), though only a weak sign is observed for a few stars
such as HD~158170 (8.4), 193370 (9.3), 180583 (9,9) which are between 
6000~K $\lesssim T_{\rm eff} \lesssim 6500$~K (given in the parentheses are 
the corresponding $v_{\rm e}\sin i$ values).
\item
Regarding lower $T_{\rm eff}$ ($< 6000$~K) stars expected have deep convective 
envelopes, emission components are observed in about $\sim 1/3$ of them (cf.
the right panel of Fig~A3) such as HD~133002 (6.6), 188650 (7.7), 141714 (7.7), 
185758 (8.4), 82210 (8.6), 164668 (9.5), 117566 (11.1), 45412 (13.1), 
198726 (14.7), 126868 (16.5), and 111812 (64.6).
\item
In summary, regarding our sample stars, emission feature in the Ca~{\sc ii} K line 
is observed in a significant fraction of $T_{\rm eff} < 6500$~K stars, though
really strong emission is seen only a few among them. 
Interestingly, however, such a feature tends to be detected mostly in not-so-rapid  
rotators of $v_{\rm e}\sin i$ around $\sim 10$~km~s$^{-1}$, while not in 
rapid rotators except for HD~111812.
\end{itemize}
Therefore, our speculation that rapidly rotating stars would have higher 
activity could not be confirmed (though this does not mean that rapid
rotators have no chromospheric activity because weak feature may possibly 
be smeared out for such broad-line cases).
We also examined whether any connection exists between the abundance anomaly
and the Ca~{\sc ii} core emission. For example, regarding the five stars 
showing anomalous [O/Fe] ratio larger than +0.5 (cf. Fig.~13k) [HD~45412 (+0.76), 
164668 (+0.80), 82210 (+0.52), 100418 (+0.62), 208110 (+0.61)], although
the former three surely show emission features (especially that of HD~82210 
is strong), such a detection is difficult (i.e., buried in noises) for 
the latter two, which means that a clear relationship can hardly be concluded. 
Accordingly, as far as this examination of Ca~{\sc ii} K line profile is 
concerned, we could not find any convincing evidence in support of the 
interpretation that large abundance scatter may have stemmed from unusual 
atmospheric condition caused by stellar chromospheric activity.  It thus seems 
premature to ascribe the reason for the large abundance dispersion to this hypothesis.

\setcounter{figure}{0}
\begin{figure*}[p]
\begin{center}
  \includegraphics[width=11cm]{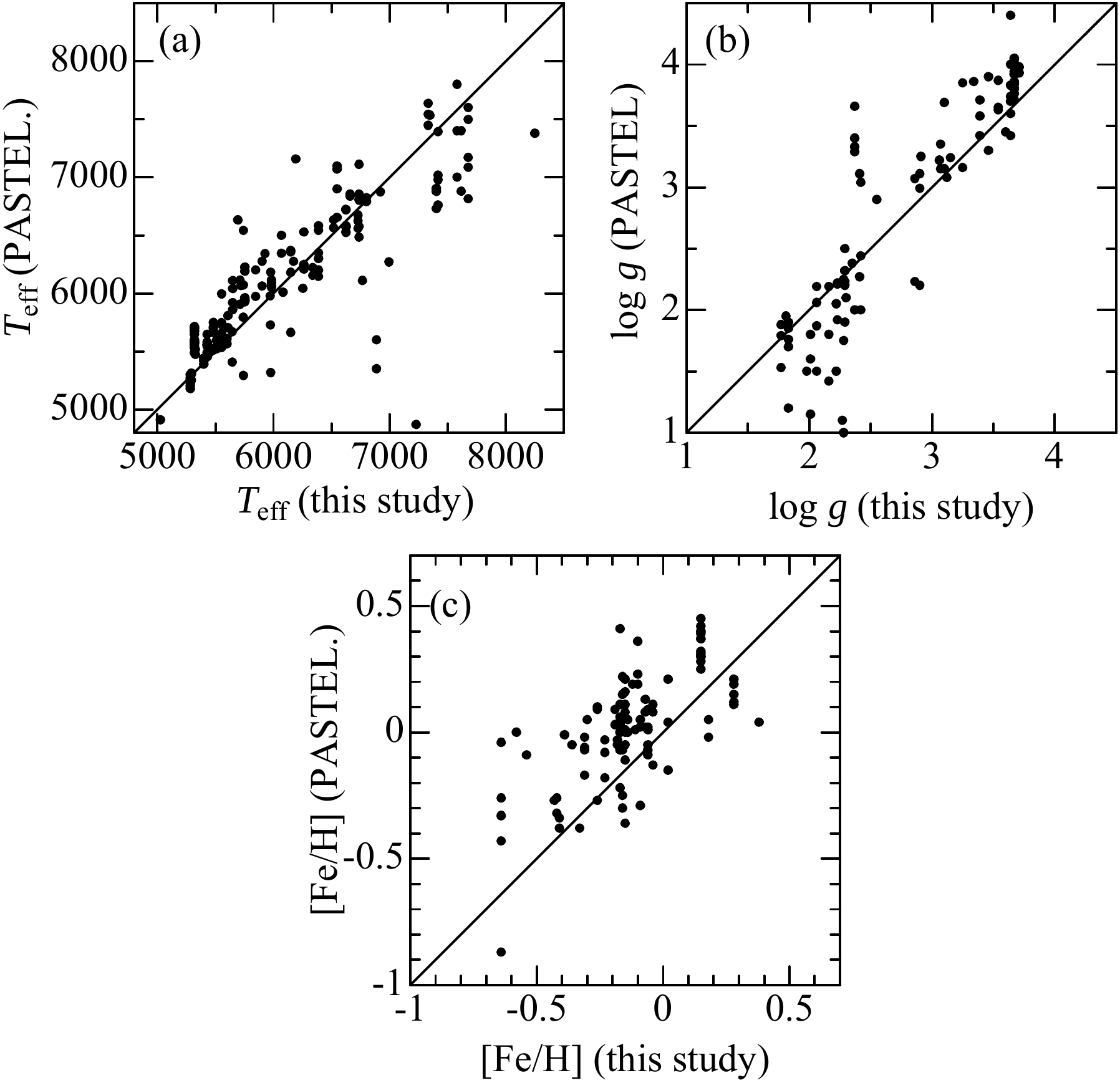}
\end{center}
\caption{
Comparison of the stellar parameters adopted in this study (taken from Paper~I) 
with those from various publications compiled in the PASTEL database 
(Soubiran et al. 2010): (a) $T_{\rm eff}$ (175 data for 57 stars in common), 
(b) $\log g$ (99 data for 40 stars in common), and (c) [Fe/H] (106 data for 
42 stars in common).
}
\label{fig:A1}
\end{figure*}

\setcounter{figure}{1}
\begin{figure*}[p]
\begin{center}
  \includegraphics[width=11cm]{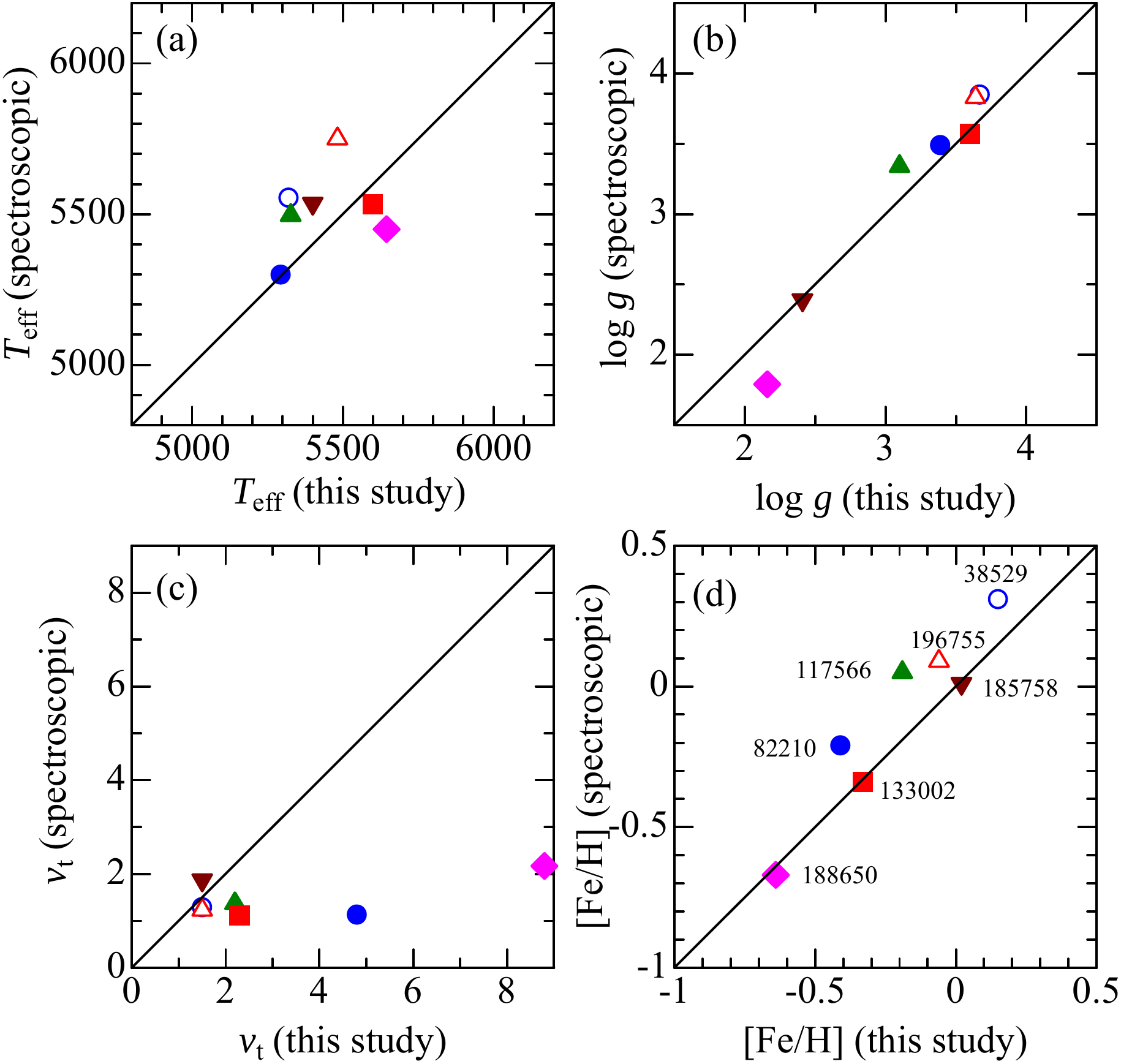}
\end{center}
\caption{
Comparison of the stellar parameters adopted in this study (taken from Paper~I) 
with those determined spectroscopically based on Fe~{\sc i} and Fe~{\sc ii} lines
by Takeda et al. (2005) (open symbols for 2 stars in common: HD 38529 and 196755) and 
Takeda et al. (2008) (filled symbols for 5 stars in common: HD 82210, 117566,
133002, 185758, and 188650). (a) $T_{\rm eff}$, (b) $\log g$, (c) $v_{\rm t}$, and (d) [Fe/H].
}
\label{fig:A2}
\end{figure*}

\setcounter{figure}{2}
\begin{figure*}[p]
\begin{center}
  \includegraphics[width=14cm]{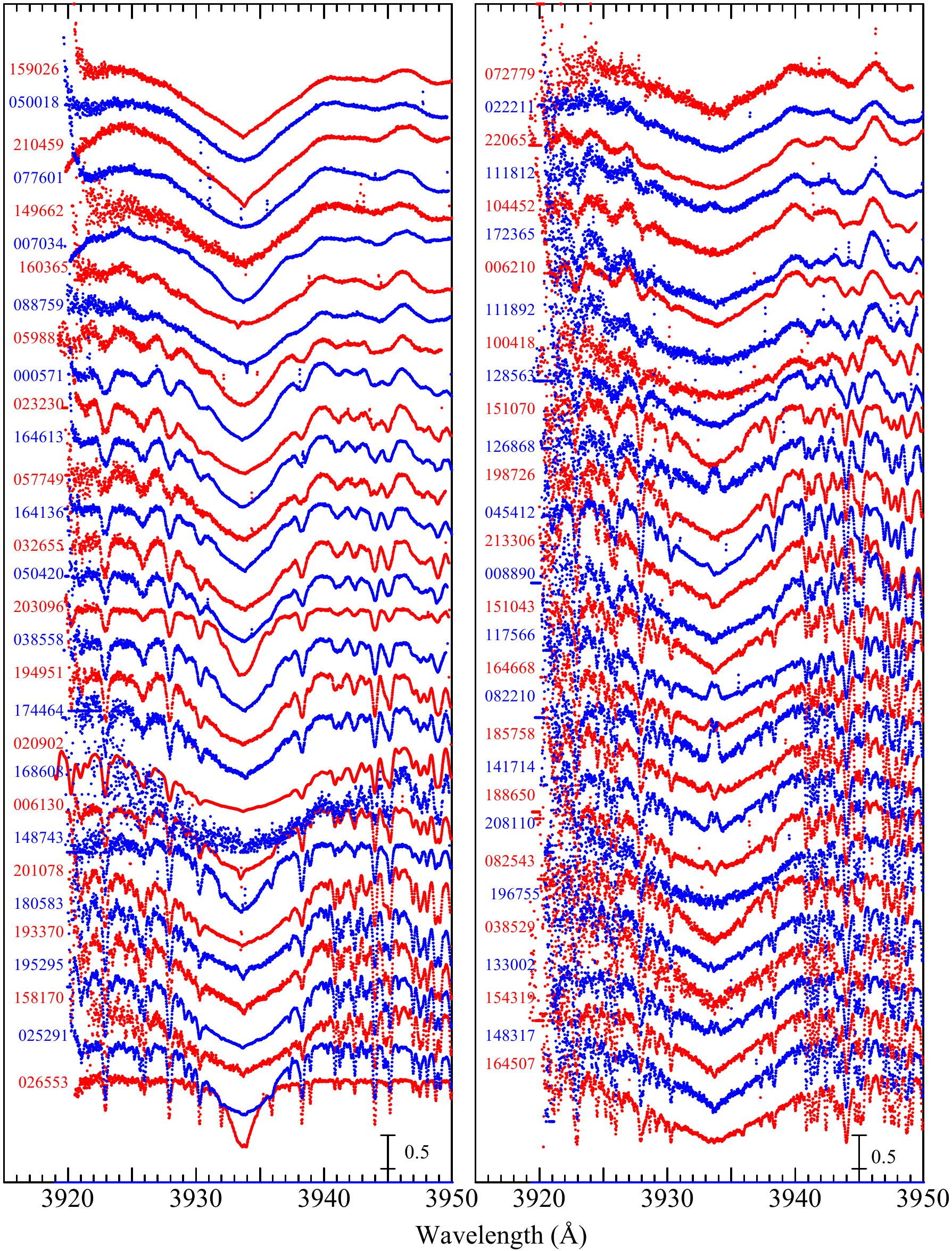}
\end{center}
\caption{
Display of the spectra in the 3920--3950~\AA\ region (comprising 
the Ca~{\sc ii} K line at 3933.663~\AA) for the 62 program stars, 
which are arranged according to the decreasing order of $v_{\rm e}\sin i$ 
as in Fig.~3 (left panel: $T_{\rm eff} > 6000$~K, right panel: 
$T_{\rm eff} < 6000$~K).
Each spectrum, tentatively normalized at $\sim$~3946~\AA, is 
vertically shifted by 0.5 relative to the adjacent one. The wavelength scale  
is adjusted to the laboratory frame by correcting the radial velocity
shifts. The HD numbers are indicated in the figure. 
}
\label{fig:A3}
\end{figure*}

\end{document}